\documentclass[english, thm-restate]{lipics-v2021}
\usepackage[utf8]{inputenc}
\usepackage{amsmath}
\usepackage{amsfonts}
\usepackage{amsthm}
\usepackage{graphicx}
\usepackage{wrapfig}
\usepackage[noend]{algpseudocode}
\usepackage{makecell}
\usepackage{xspace}
\usepackage{caption}
\usepackage{imakeidx}
\usepackage{thmtools}
\usepackage{makecell}
\usepackage[numbers,sort]{natbib}
\usepackage{multirow}

\usepackage{hyperref}

\newcommand{\etal}{\textit{et al.}\xspace}
\newcommand{\tor}{\!\to\!}

\newtheorem{oracles}[theorem]{Oracles}

\newcommand{\R}{\ensuremath{\mathbb R}}
\newcommand{\eps}{\ensuremath{\varepsilon}}

\newcommand{\X}{\mathcal{X}\xspace}

\renewcommand{\d}{\ensuremath{d^\circ}}
\newcommand{\frechet}{Fr\'echet\xspace}
\newcommand{\FD}{\ensuremath{{D_{\mathcal{F}}} }}
\newcommand{\FDd}{\ensuremath{{{D_{\mathbb{F}}} }}}
\newcommand{\HD}{\ensuremath{{D_{\mathcal{H}}} }}
\newcommand{\dist}[1]{\mathcal{D}_\X^{#1}}

\title{Data Structures for Approximate Fréchet Distance for Realistic Curves}
\titlerunning{Data Structures for Approximate Fréchet Distance for Realistic Curves}

\author{Ivor van der Hoog}{DTU Compute, Technical University of Denmark, Denmark}{idjva@dtu.dk}{0009-0006-2624-0231}{Supported by Independent Research Fund Denmark grant 2020-2023 (9131-00044B) ``Dynamic Network Analysis''.}

\author{Eva Rotenberg}{DTU Compute, Technical University of Denmark, Denmark}{erot@dtu.dk}{0000-0001-5853-7909}{Partially supported by Independent Research Fund Denmark grant 2020-2023 (9131-00044B) ``Dynamic Network Analysis'' and the Carlsberg Young Researcher Fellowship CF21-0302 ``Graph Algorithms with Geometric Applications''.}

\author{Sampson Wong}{Department of Computer Science, University of Copenhagen, Denmark}{sampson.wong123@gmail.com}{0000-0003-3803-3804}{}

\nolinenumbers

\authorrunning{Ivor van der Hoog, Eva Rotenberg, Sampson Wong}

\Copyright{Ivor van der Hoog, Eva Rotenberg, Sampson Wong} 


\date{}
\begin{document}

\maketitle

\begin{abstract}
The Fr\'{e}chet distance is a popular distance measure between curves $P$ and $Q$. Conditional lower bounds prohibit $(1 + \eps)$-approximate Fr\'{e}chet distance computations in strongly subquadratic time, even when preprocessing $P$ using any polynomial amount of time and space. As a consequence, the Fr\'echet distance has been studied under \emph{realistic} input assumptions, for example, assuming both curves are \emph{$c$-packed}.

In this paper, we study $c$-packed curves in Euclidean space $\mathbb R^d$ and in general geodesic metrics $\X$. 
In $\mathbb R^d$, we provide a nearly-linear time static algorithm for computing the $(1+\varepsilon)$-approximate continuous Fr\'echet distance between $c$-packed curves. Our algorithm has a linear dependence on the dimension~$d$, as opposed to previous algorithms which have an exponential dependence on~$d$.
 
In general geodesic metric spaces $\mathcal X$, little was previously known. We provide the first data structure, and thereby the first algorithm, under this model. Given a $c$-packed input curve $P$ with $n$ vertices, we preprocess it in $O(n \log n)$ time, so that given a query containing a constant $\varepsilon$ and a curve $Q$ with $m$ vertices, we can return a $(1+\varepsilon)$-approximation of the discrete Fr\'echet distance between $P$ and $Q$ in time polylogarithmic in $n$ and linear in $m$, $1/\varepsilon$, and the realism parameter~$c$.

Finally, we show several extensions to our data structure; to support dynamic extend/truncate updates on $P$, to answer map matching queries, and to answer Hausdorff distance queries.
\end{abstract}

\section{Introduction}

The \frechet distance is a popular metric for measuring
the similarity between (polygonal) curves $P$ and $Q$.
We assume that $P$ has $n$ vertices and $Q$ has $m$ vertices and that they reside in some geodesic metric space $\X$.  
The \frechet distance is often intuitively defined through the following metaphor: suppose that we have two curves that are traversed by a person and their dog. Consider the length of their connecting leash, measured over the metric $\X$. What is the minimum length of the connecting leash over all possible traversals by the person and the dog?
The \frechet distance
has many applications; in particular in the analysis and
visualization of movement data~\cite{ buchin2017clustering, buchin2020group,  konzack2017visual, xie2017distributed}.
It is a versatile measure that can be used for a variety of objects, such as handwriting \cite{sriraghavendra2007frechet},  coastlines~\cite{mascret2006coastline}, outlines of shapes in geographic information systems~\cite{devogele2002new}, trajectories of moving objects, such as vehicles, animals or sports players~\cite{acmsurvey20, su2020survey, brakatsoulas2005map, buchin2020group}, air traffic~\cite{ bombelli2017strategic} and protein structures~\cite{jiang2008protein}. 

 Alt and Godau~\cite{alt1995computing} compute the continuous \frechet distance in $\R^2$ under the $L_2$ metric in $O(mn \log (n + m))$ time. This was later improved by Buchin~\etal~\cite{buchin2017four} to $O(nm (\log \log  nm )^2)$ time. Eiter and Manila~\cite{eitermannila94} showed how to compute the discrete \frechet distance in $\R^2$ in $O(nm)$ time, which was later improved by Agarwal~\etal~\cite{agarwal2014computing} to $O(nm (\log \log  nm ) / \log nm )$ time.
Typically, the quadratic $O(nm)$ running time is considered costly. Bringmann~\cite{bringmann2014walking} showed that, conditioned on the Strong Exponential Time Hypothesis (SETH), one cannot compute a $(1 + \eps)$-approximation of the continuous \frechet distance between curves in $\R^2$ under the $L_1, L_2$ or $L_\infty$ metric faster than $\Omega( (nm)^{1 - \delta})$ time for any $\delta > 0$. 
This lower bound was extended by Buchin, Ophelders and Speckmann~\cite{buchin2019seth} to intersecting curves in $\mathbb{R}^1$. Driemel, van der Hoog and Rotenberg~\cite{driemel2023discrete} extended the lower bound to paths $P$ and $Q$ in a weighted planar graph under the shortest path metric.


\subparagraph{Well-behaved curves.}
Previous works have circumvented lower bounds by assuming that \emph{both} curves come from a well-behaved class.
A curve $P$ in a geodesic metric space $\X$ is any sequence of points where consecutive points are connected by their shortest path in $\X$. For a ball $B$ in $\X$, let $P \cap B$ denote all (maximal) segments of $P$ contained in $B$. 
A curve $P$ is: 
\begin{itemize}
    \item $\kappa$-\textbf{straight} (by Alt, Knauer and Wenk~\cite{alt2004comparison}) if for every $i, j$ the length of the subcurve from $p_i$ to $p_j$ is $\ell(P[i, j]) \leq \kappa \cdot d(p_i, p_j)$,
    \item $c$-\textbf{packed} (by Driemel, Har-Peled and Wenk~\cite{DriemelHW12}) if for every ball $B$ in the geodesic metric space $\X$ with radius $r$: the length  $\ell(P \cap B) \leq c \cdot r$.
    \item $\phi$-low-\textbf{dense} (by van der Stappen~\cite{DBLP:books/daglib/vanderstappen}; see also \cite{devogele2002new,DriemelHW12,DBLP:journals/ipl/SchwarzkopfV96}) if for every ball $B$ in $\X$ with radius $r$, there exist at most $\phi$ edges of length~$r$ intersecting~$B$.
    \item \textbf{backbone} (by Aronov et al.~\cite{DBLP:conf/esa/AronovHKWW06}) if consecutive vertices have distance between~$c_1$ and~$c_2$ for some constants $c_1, c_2$, and if non-consecutive vertices have distance at least 1.
\end{itemize}
\noindent
Any $c$-straight curve is also $O(c)$-packed.
Parametrized by $\varepsilon$, $\phi \in O(1)$, $c$ and $\kappa = O(c)$, Driemel, Har-Peled and Wenk~\cite{DriemelHW12} compute a $(1+\varepsilon)$-approximation of the continuous Fr\'echet distance between a pair of realistic curves in $\R^d$ under the $L_1, L_2, L_\infty$ metric for constant $d$ in $O(\frac{ c (n+m)}{\eps} + c (n+m) \log n)$ time. 
Their result for $c$-packed and $c$-straight curves was improved by Bringmann and K\"{u}nnemann~\cite{bringmann2017improved} to $O( \frac{c (n+m) }{\sqrt{\eps}} \log \eps^{-1} + c (n+m) \log n )$, which matches the conditional lower bound for $c$-packed curves. In particular, Bringmann~\cite{bringmann2014walking} showed that under SETH, for dimension $d \geq 5$, there is no $O( (c(n+m)/\sqrt{\varepsilon})^{1 - \delta})$ time algorithm for computing the Fr\'echet distance between $c$-packed curves for any $\delta > 0$. 
Realistic input assumptions have been applied to other geometric problems, e.g. for robotic navigation in $\phi$-low-dense environments~\cite{DBLP:books/daglib/vanderstappen}, and map matching of $\phi$-low-dense graphs~\cite{DBLP:conf/alenex/ChenDGNW11} or $c$-packed graphs~\cite{gudmundsson:mapmatching}.

\subparagraph{Deciding versus computing.}
We make a distinction between two problem variants: the decision variant, the optimisation variant. 
For the decision variant, we are given a value $\rho$ and two curves $P$ and $Q$ and we ask whether the \frechet distance $\FD(P, Q) \leq \rho$. This variant often solved through navigating an $n$ by $m$ `free space diagram'.
In the optimization variant, the goal is to output the Fr\'echet distance $\FD(P, Q)$. 
To convert any decision algorithm into a optimization  algorithm, two techniques are commonly used. 
The first is binary search over what we will call TADD$(P, Q)$:  

\begin{definition}
    Given two sets of points $P$, $Q$ in a geodesic metric space $\X$, we define a Two-Approximate Distance Decomposition of $P$ denoted by \textbf{TADD($P, Q$)} as a set of reals $T_{PQ}$ where for every pair $(p_i, q_j) \in P \times Q$ there exist $a, b \in T_{PQ}$ with $a \leq d(p_i, q_j) \leq b \leq 2a$. 
\end{definition}

Essentially a TADD is a two-approximation of the set of all pairwise distances in $P \times Q$ and it can be used to determine, approximately, the (Fr\'echet) distance values for when the simplification of the input curve changes, or when the reachability of the free space matrix changes. 
It is known how to compute a TADD from a Well-Separated Pair Decomposition (WSPD) in time linear in the size of the WSPD~\cite[Lemma 3.8]{DriemelHW12}.
A downside of this approach~\cite{DriemelHW12} is that, it is only known how to compute a WSPD for doubling metrics~\cite{xu2005well}. Moreover, for non-constant (doubling) dimensions $d$, computing the WSPD (and therefore the TADD) takes  $O(2^d n + d n \log n)$ time~\cite{har2011geometric,xu2005well}, which dominates the running time. 

The second technique,
deployed when for example TADDs cannot be computed, 
is parametric search~\cite{DBLP:journals/jacm/Megiddo83}. For decision variants that have a sublinear running or query time of $T$, the running time of parametric search is commonly $O(T^2)$~\cite{van2002parametric, gudmundssonfrechet2020}.

\subparagraph{Data structures for \frechet distance.}
An interesting question is whether we can store $P$ in a data structure, for efficient (approximate) \frechet distance queries for any query $Q$. This topic received considerable attention throughout the
years~\cite{DriemelPS19, gudmundssonfrechet2020, FiltserF21, de2017data, driemel2013jaywalking, buchin2022efficient,  gudmundsson:mapmatching}.
A related field is nearest neighbor data structures under the \frechet distance metric~\cite{DBLP:conf/soda/BringmannDNP22, driemelpsarros2020, de2013fast, aronov2019:nearest_neigbhbor_queries_planar_curves, filtser2020approximateICALP}.
Recently, Gudmundsson, Seybold and Wong~\cite{gudmundsson:mapmatching} answer this question negatively for arbitrary curves in $\R^2$: showing that even with polynomial preprocessing space and time, we cannot preprocess a curve $P$ 
to decide the continuous \frechet distance between $P$ and a query curve $Q$ in $\Omega( (nm)^{1 - \delta})$ time for any $\delta > 0$. 
Surprisingly, even in very restricted settings data structure results are difficult to obtain. De Berg~\etal~\cite{de2017data} present an $O(n^2)$ size
data structure that restricts the orientation of the query segment to be horizontal. Queries are supported in
$O(\log^2 n)$ time, and even subcurve queries are allowed (in that case, using
$O(n^2\log^2 n)$ space). At ESA 2022, Buchin~\etal~\cite{buchin2022efficient} improve these result to using only $O(n\log^2 n)$ space, where queries take $O(\log n)$ time.
For arbitrary query segments, they present an $O(n^{4+\delta})$ size data structure that supports (subcurve) queries to arbitrary segments in $O(\log^4 n)$ time. 
Gudmundsson~\etal~\cite{gudmundssonfrechet2020} extend de Berg~\etal's~\cite{de2017data} data structure to handle subcurve queries, and to handle queries where the horizontal query segment is translated in order to minimize its Fr\'echet distance. 
Driemel and Har-Peled~\cite{driemel2013jaywalking} create a data structure to store any curve in $\R^d$ for constant $d$.
They preprocess $P$ in $O(n \log^3 n)$ time and $O(n \log n)$ space. For any query $(Q, \eps, i, j)$ they can create a $(3 + \eps)$-approximation of $\FDd(P[i, j], Q)$ in $O(m^2 \log n \log(m \log n))$ time.

We state existing data structures for the discrete Fr\'echet distance. Driemel, Psarros and Schmidt~\cite{DriemelPS19} fix $\eps$ and an upper bound $M$  beforehand where for all queries $Q$, they demand that $|Q| \leq M$. 
They store any curve $P$ in $\R^d$ for constant $d$ using $O( (M \log \frac{1}{\eps})^M )$ space and preprocessing, to answer $(1+\eps)$-approximate \frechet distance queries in $O(m^2 + \log \frac{1}{\eps})$ time. 
Filtser~\cite{Filtser18} gives the corresponding data structure for the discrete \frechet distance.
At SODA 2022, Filtser and Filtser~\cite{FiltserF21} study the same setting: storing $P$ in $O(  \left( \frac{1}{\eps} \right)^{dM} \log \frac{1}{\eps})$ space, to answer $(1+\eps)$-approximate \frechet distance queries in $\Tilde{O}(m \cdot d)$ time.

\subparagraph{Contributions.}

\begin{table}[]
{\tiny
\hspace{-1.75cm}
\renewcommand{\arraystretch}{2.5} 
    \begin{tabular}{|c|c|c|c|c|c|c|c|}
    \hline
     \multirow{2}{*}{Domain} &  \multicolumn{3}{c|}{Previous result} & \multicolumn{3}{c|}{Our result}\\
        \cline{2-7}
          & Preprocess & Query time & Ref. & Preprocess  & Query time & Ref.\\
    \hline
        \makecell[c]{$\R^d$ \\
        $L_p$ metric } & static & \makecell[c]{$\tilde{O}(2^d n + d \frac{c (n+m)}{\sqrt{\eps}}$ \\ $+ d^2 \cdot c(n+m))$} & \cite{bringmann2017improved}  & static & $ \tilde{O}( d \frac{c (n + m)}{\eps}) $ & Thm.~\ref{thm:TADD}
           \\
                     \hline
           \multicolumn{7}{|c|}{\textbf{Discrete \frechet distance only:}} \\
         \hline
           \makecell[c]{$\R^d$ \\ $|Q| \leq M$ } &  $ \tilde{O}( (\frac{1}{\eps})^{d M} )$ & $\tilde{O}(m d)$ & \cite{FiltserF21} & $O(n \log n)$ & $ \tilde{O}( d \frac{cm}{\eps} \log n) $ & Cor.~\ref{cor:frechetvalue}
           \\
                  \hline
          \makecell[c]{ Planar $G = (V, E)$ } & static & $\tilde{O} (|V|^{1 + o(1)} + \frac{c(n+m)}{\eps}) $ & \cite{driemel2023discrete} & $\tilde{O}(|V|^{1 + o(1)})$ & $ \tilde{O}( \frac{cm}{\eps} \log n) $ & Cor.~\ref{cor:frechetvalue}
           \\
                  \hline
          \makecell[c]{ Graph $G = (V, E)$ } & static & \makecell[c]{$\tilde{O}(|V|^{1 + o(1)} + |E| \log |E|$ \\ $+ \frac{c(n+m)}{\eps} )$} & \cite{driemel2023discrete} & $\tilde{O}(|V|^{1+ o(1)})$ & $ \tilde{O}( \frac{cm}{\eps} \log n) $ & Cor.~\ref{cor:frechetvalue}\\
           
      \hline
         \makecell[c]{ Simple  Polygon $P$ }  & static & $O(nm \log (n+m)) $  &  \cite{alt1995computing} & $\tilde{O}(|P| + n)$ & $\tilde{O}(\frac{cm}{\eps} \log |P|)$ & Cor.~\ref{cor:frechetvalue}  \\
    \hline 
        \makecell[c]{ Any geodesic $\mathcal{X}$ \\ with $(1+\varepsilon)$-oracle} & static & $O(T_\eps \cdot nm \log n ) $  & \cite{eitermannila94} & ${O}(n \log n)$ & $\tilde{O}(T_{\eps} \frac{cm}{\eps}\log n)$  & Thm.~\ref{thm:frechet_value}  \\         
         \hline
         \makecell[c]{Map Matching \\
         $G = (V, E)$} & $\tilde{O}(c^2 \eps^{-4} n^2)$ & \makecell[c]{$O(m \log m \cdot $ \\ $ (\log^4 n + c^4 \eps^{-8} \log^2 n ))$} & \cite{gudmundsson:mapmatching} & $\tilde{O}(c^2 \eps^{-4} n^2)$  &  \makecell[c]{$O(m (\log n + \log \eps^{-1}) \cdot$ \\ $(\log^2 n + c^2 \eps^{-4}\log n))$} & Cor.~\ref{cor:mapmatching}\\
   \hline
    \end{tabular}
    }
    \caption{
    Our results for computing a $(1+\eps)$-approximation of the \frechet distance between $P$ and $Q$. 
    All settings assume that $P$ is a realistic curve, except for \cite{FiltserF21} who assume an upper bound on $|Q|$.   
    $T_\varepsilon$ denotes the query time of a $(1+\varepsilon)$-approximate oracle.
    The tilde hides lower order factors in terms of $n$, $m$ and~$\varepsilon$.   
    Under the continuous Fr\'{e}chet distance, we require that $Q$ is also realistic. \vspace{-0.5cm}
    }
    \label{tab:results}
\end{table}
We provide four contributions.


\emph{(1) A $1$-TADD technique.} A crucial step in computing the Fr\'echet distance is to turn a decision algorithm into an optimization algorithm. TADD($P, Q$) is commonly used when approximating the Fr\'echet distance. Our 1-TADD technique shows a new argument where we map $P$ and $Q$ to curves in $\Lambda \subset \R^1$ and compute only TADD($\Lambda, \Lambda$) in $O(n \log n)$ time.

In Euclidean $\mathbb R^d$ this allows us  to approximate the discrete and continuous Fr\'echet distances in time that is linear in $d$, whereas previous approaches required an exponential dependence on~$d$. In general geodesic metric spaces~$\mathcal X$, our 1-TADD technique allows us to approximate the discrete Fr\'echet distance when  TADD$(P, Q)$ cannot be efficiently computed.

\emph{(2) Allowing approximate oracles under the discrete \frechet distance.}
Many ambient spaces (e.g., Euclidean spaces under floating point arithmetic, and $\X$ as a weighted graph under the shortest path metric.) do not allow for efficient exact distance computations.  
Thus, we revisit and simplify the argument by Driemel, Har-Peled and Wenk~\cite{DriemelHW12}. We assume access to a $(1+\alpha)$-approximate distance oracle with $T_\alpha$ query time. 
We generalize the previous argument to approximate the discrete \frechet distance between two curves in any geodesic metric space with approximate distance oracles. For contributions \emph{(1)} and \emph{(2)}, we do not require the curves~$P$ and~$Q$ to be $c$-packed.

\emph{ (3) A data structure under the discrete \frechet distance.}
Under the discrete \frechet distance, we show how to store a $c$-packed or $c$-straight curve $P$ with $n$ vertices in \emph{any} geodesic ambient space $\X$. 
Our solution uses $O(n)$ space and $O(n \log n)$ preprocessing time.
For \emph{any} query curve $Q$, any $0 < \eps < 1$, and any subcurve $P^*$ of $P$, we can compute a $(1+\eps)$-approximation the discrete Fr\'echet distance $\FDd(P^*, Q)$ using $O( \frac{c \cdot m}{\eps} \log n (T + \log \frac{c \cdot m}{\eps} + \log n))$ time. 
Here, $T$ is the time required to perform a distance query in the ambient space (e.g., $O(\log n)$ for geodesic distances in a polygon). All times are deterministic and worst-case. This is the \frechet distance first data structure for realistic curves that avoids spending query time linear in $n$. Our solution improves various recent results~\cite{bringmann2017improved, driemel2023discrete, FiltserF21, gudmundsson:mapmatching} (see Table~\ref{tab:results}). 

\emph{(4) Extensions.}
In Section~\ref{sec:dynamic}, we modify our data structure to support updates that truncate the curve~$P$, or extend $P$, we.  In Section~\ref{app:mapmatching}, we apply our algorithmic skeleton to map matching queries, and w. In Section~\ref{app:hausdorff}, we study Hausdorff distance queries.



\section{Preliminaries}
\label{sec:prelims}

Let $\X$ denote some geodesic metric space (e.g., $\X$ is some weighted graph). 
For any $a, b \in \X$ we denote by $d(a, b)$ their distance in $\X$. A \emph{curve} $P$ in $\X$ is any ordered sequence of points in $\X$, where consecutive points are connected by their shortest path in $\X$. 
We refer to such points as \emph{vertices}. 
For any curve $P$ with $n$ vertices, for any integers $i, j \in [n]$ with $i < j$ we denote by $P[i, j]$ the subcurve from $p_i$ to $p_j$. 
We denote by $|P[i, j]| = (j-i+1)$ the size of the subcurve and by $\ell(P[i, j]) := \sum_{k = i}^{j-1} d(p_k, p_{k+1})$ its length. 
We receive as preprocessing input a curve $P$ where for each pair $(p_i, p_{i+1})$ we are given $d(p_i, p_{i+1})$. 

\subparagraph{Distance and distance oracles.}
Throughout this paper, we assume that for any $\alpha > 0$ we have access to some $(1+\alpha)$-approximate distance oracle. 
This is a data structure $\dist{\alpha}$ that for any two $a,b \in \X$ can report a value $\d(a, b) \in \left[ (1-\alpha) d(a, b), (1 + \alpha) d(a, b) \right]$ in $O(T_\alpha)$ time.
To distinguish between inaccuracy as a result of our algorithm and as a result of our oracle, we refer to $\d(a, b)$ as the \emph{perceived value} (as opposed to an approximate value).  

\subparagraph{Discrete \frechet distance. }
Given two curves $P$ and $Q$ in $\X$, we denote by $[n] \times [m]$ the  $n$ by $m$ integer lattice.
We say that an ordered sequence $F$ of points in $[n] \times [m]$ is a \emph{discrete walk} if for every consecutive pair $(i, j), (k, l) \in F$, 
we have $k\in\{i-1,i,i+1\}$ and $l\in \{j-1,j,j+1\}$. It is furthermore \emph{$xy$-monotone} when we restrict to $k\in \{i,i+1\}$ and $l\in\{j,j+1\}$.
Let $F$ be a discrete walk from $(1, 1)$ to $(n, m)$. 
The \emph{cost} of $F$ is the maximum over $(i, j) \in F$ of $d(p_i, q_j)$. 
The discrete \frechet distance is the minimum over all $xy$-monotone walks $F$ from $(1, 1)$ to $(n, m)$ of its associated cost:
$
\FDd (P, Q) := \min_{F }  \textnormal{cost}(F) = \min_{F } \max_{(i, j) \in F} d(p_i, q_j).
$

\noindent
In this paper we, given a $(1+\alpha)$-approximate distance oracle, define the \emph{perceived} discrete \frechet distance as ${\FDd^\circ}$, obtained by replacing in the above definition  $d(p_i, q_j)$ by $\d(p_i, q_j)$.

\subparagraph{Free space matrix (FSM).}
The FSM for a fixed $\rho^* \geq 0$ is a $|P| \times |Q|$,  $(0,1)$-matrix where the cell $(i, j)$ is zero if and only if the distance between the $i$'th point in $P$ and the $j$'th point in $Q$ is at most $\rho^*$.  Per definition, ${\FDd}(P, Q) \leq \rho^*$ if and only if there exists an $xy$-monotone discrete walk $F$ from $(1, 1)$ to $(n, m)$ where for all $(i, j) \in F$: the cell $(i, j)$ is zero.

\subparagraph{Continuous Fr\'echet distance and Free Space Diagram (FSD)} We define the continuous Fr\'echet distance in a geodesic metric space. Given a curve $P$, we consider $P$ as a {continuous} function mapping at time $t \in [0,1]$ to a point $P(t)$ in $\X$. 
The continuous Fr\'echet distance is $\FD (P,Q) := \inf_{\alpha,\beta} \max_{t \in [0,1]} d(P(\alpha(t)),Q(\beta(t))$, where $\alpha,\beta : [0, 1] \to [0,1]$ are non-decreasing surjections. For a fixed $\rho^*$, we can define the Free Space Diagram of $(P, Q, \rho^*)$ to be a $[0,1] \times [0,1]$, (0,1)-matrix where the point $(t,t')$ is zero if and only if the distance between $P(t)$ and $Q(t')$ is at most $\rho^*$. The diagram consists of $nm$ cells corresponding to all pairs of edges of $P$ and $Q$. A cell is \emph{reachable} if there exists an $xy$-monotone curve from $(0, 0)$ to a point in the cell where all points $(t, t')$ on the curve are zero. 
The continuous Fr\'echet distance is at most $\rho^*$ if and only if there exists an $xy$-monotone curve from $(0,0)$ to $(1,1)$ where all points $(t,t')$ on the curve are zero.

\subparagraph{Defining discrete queries.}
Our data structure input is a curve $P = (p_1, p_2, \ldots, p_n)$.
The number of vertices $m$ of $Q$ is part of the query input and may vary. 
Let $\FDd(P, Q)$ denote the discrete \frechet distance between $P$ and $Q$. We distinguish between four types of (approximate) queries. The input pararmeters are given at query time:
\begin{itemize}
    \item \textbf{A-decision$(Q, \eps, \rho)$:}  for $\rho \geq 0$ and $0 < \eps < 1$ outputs a Boolean concluding either $\FDd(P, Q) > \rho$, or $\FDd(P, Q) \leq (1 + \eps) \rho$ (these two options are not mutually exclusive).
    \item \textbf{A-value$(Q, \eps)$:} for  $0 < \eps < 1$ outputs a value in $[(1-\eps) \FDd(P, Q), (1+\eps) \FDd(P, Q)]$.
    \item \textbf{Subcurve-decision($Q, \eps, \rho, i, j$):} for $\rho \geq 0$ and $0 < \eps < 1$ outputs a Boolean concluding either $\FDd(P[i, j], Q) > \rho$, or $\FDd(P[i, j], Q) \leq (1 + \eps) \rho$.
    \item \textbf{Subcurve-value($Q$, $\eps$, $i$, $j$):} for  $0 < \eps < 1$ outputs a value in 
    
    $[(1-\eps) \FDd(P[i,j], Q), (1+\eps) \FDd(P[i, j], Q)]$.
\end{itemize}
We want a solution that is efficient in time and space, where time and space is measured in units of $\eps, n, m, \rho$ and the distance oracle query time $T_\alpha$.

\subparagraph{Previous works: $\mu$-simplifications.}

Driemel, Har-Peled and Wenk~\cite{DriemelHW12} , for a parameter $\mu \in \R$, construct a curve $P^\mu$ as follows. Start with the initial vertex $p_1$, and set this as the current vertex $p_i$. Next, scan the polygonal curve to find the first vertex $p_j$ such that $d(p_i, p_j) > \mu$. Add $p_j$ to $P^\mu$, and set $p_j$ as the current vertex. Continue this process until we reach the end of the curve. Finally, add the last vertex $p_n$ to $P^\mu$.  Driemel, Har-Peled and Wenk~\cite{DriemelHW12} observe any $\mu$-simplified curve $P^\mu$ can be computed in linear time and $\FD(P^\mu, P) \leq \mu$.
This leads to the following approximate decision algorithm:

\begin{enumerate}
    \item Given $P, \eps, Q$ and $\rho$, construct $P^{\frac{\eps \rho}{4}}$ and $Q^{\frac{\eps \rho}{4}}$ in $O(n + m)$ time.
    \item Denote by $X$ the reachable cells in the FSD of $(P^{\frac{\eps \rho}{4}}, Q^{\frac{\eps \rho}{4}}, \rho^* = (1 + \eps/2)\rho)$.
    \item Iterating over $X$, doing $O(|X|)$ distance computations, test if $\FD(P^{\frac{\eps \rho}{4}}, Q^{\frac{\eps \rho}{4}}) \leq \rho^*$.
    \begin{itemize}
        \item They prove that: if yes then $\FD(P, Q) \leq (1 + \eps)\rho$. If no then
        $\FD(P, Q) > \rho$.
        \item If $P$ and~$Q$ are $c$-packed, they upper bound $|X|$ by $O(\frac{c(n + m)}{\eps})$.
    \end{itemize}
\end{enumerate}

\noindent
This scheme is broadly applicable to various domains, see~\cite{driemel2023discrete, conradi2024revisiting}. 
In this paper, we apply this technique to answer value queries at the cost of a factor $O(\log n + \log \eps^{-1})$. 
Under the discrete \frechet distance, we extend the analysis to work with approximate distance oracles. Finally, we show a data structure to execute step 3 in time independent of $|P| = n$. We also show that under the discrete Fr\'echet distance, it suffices to assume that only $P$ is $c$-packed.

\subsection{Results.}

\emph{(1) A $1$-TADD technique for the \frechet distance.}
For any $\mu > 0$, we denote by $P^\mu$ and $Q^\mu$ their $\mu$-simplified curves according to our new definition of $\mu$-simplification. 
We show in Section~\ref{sec:1TADD} that our new definition allows us to efficiently transform existing decision algorithms into approximation algorithms in $\mathbb{R}^d$~\cite{DriemelHW12}.
We assume access to exact $O(d)$-time distance oracle in $\mathbb R^d$ and prove:

\begin{restatable}{theoremrestate}{TADD}
  \label{thm:TADD}
    We can preprocess a pair of $c$-packed curves $(P,Q)$ in $\mathbb{R}^d$ under any $L_p$ metric with $|P| =  n \geq |Q| = m$ in $O( n \log n)$ time s.t.: given any $\eps$ and an exact distance oracle, we can compute a $(1+\eps)$-approximation of $\FD(P, Q)$ in $O( d \frac{c(n+m)}{\eps} \cdot (\log n + \log \eps^{-1}))$ time. 
\end{restatable}

\emph{(2) Allowing approximate oracles under the discrete \frechet distance. }
In Section~\ref{sec:approximate}, we show that (for computing the discrete \frechet distance)  it suffices to have access to a $(1 + \alpha)$-approximate distance oracle. This will enable us to approximate $\FDd(P, Q)$ in ambient spaces such as planar graphs and simple polygons.
Formally, we show:

\begin{restatable}{lemmarestate}{decisionanswer}
\label{lem:decision_answer}
For any $\rho > 0$ and $0 < \eps < 1$, choose $\rho^* = (1 + \frac 1 2 \eps)\rho$ and $\mu \leq \frac{1}{6} \eps \rho$.
Let $\X$ be any geodesic metric space and $\dist{\eps / 6}$ be a $(1 + \frac 1 6 \eps)$-approximate distance oracle. 
For any curve $P = (p_1, \ldots, p_n)$ in $\X$ and any curve $Q = (q_1, \ldots, q_m)$ in $\X$:
\begin{itemize}
    \item If for the discrete \frechet distance, $\FDd^\circ(P^\mu, Q) \leq \rho^*$ then $\FDd(P, Q) \leq (1 + \eps)\rho$.
    \item If for the discrete \frechet distance, $\FDd^\circ(P^\mu, Q) > \rho^*$ then $\FDd(P, Q) > \rho$.
\end{itemize}
\end{restatable}

\noindent
We note that for ambient spaces such as planar graphs and simple polygons, there is no clear way to define a continuous $\mu$-simplification (or even continuous \frechet distance).

\emph{(3) An efficient data structure under the discrete \frechet distance. }
Finally, in Section~\ref{sec:frechet}, we study computing the discrete \frechet distance in a data structure setting.
We show that under the discrete \frechet distance, it suffices to assume that only $P$ is $c$-packed or $c$-straight. Moreover,  we can store $P$ in a data structure such that we can answer approximate Fr\'echet value queries $\FDd(P, Q)$ in time linear in $m$ and polylogarithmic in~$n$:

\begin{restatable}{theorem}{frechetdecision}
\label{thm:frechetdecision}
Let $\X$ be any geodesic space and $\dist{\alpha}$ be a $(1 + \alpha)$-approximate distance oracle with $O(T_\alpha)$ query time. 
Let $P = (p_1, \ldots, p_n)$ be any $c$-packed curve in $\X$. 
We can store $P$ using $O(n)$ space and preprocessing, such that for any curve $Q = (q_1, \ldots, q_m)$ in $\X$ and any $\rho > 0$ and $0 < \eps < 1$, we can answer A-decision$(Q, \eps, \rho)$ for the discrete \frechet distance in: 
\[O \left( \frac{c \cdot m}{\eps} \cdot \left( T_{\eps / 6} + \log n \right) \right) \textnormal{ time.}\]
\end{restatable}

We may apply the proof of Theorem~\ref{thm:TADD} to Theorem~\ref{thm:frechetdecision} to answer A-value$(Q, \eps)$ in any geodesic metric space by increasing the running time by a factor $O(\log n + \log \eps^{-1})$. However, under the discrete \frechet distance we can be more efficient:

\begin{restatable}{theoremrestate}{frechetvalue}
\label{thm:frechet_value}
Let $\X$ be a geodesic metric space and $\dist{\alpha}$ be a $(1 + \alpha)$-approximate distance with $O(T_\alpha)$ query time. 
Let $P = (p_1, \ldots, p_n)$ be any $c$-packed curve in $\X$. 
We can store $P$ using $O(n)$ space and $O(n \log n)$ preprocessing time, such that for any curve $Q = (q_1, \ldots, q_m)$ in $\X$ and any $0 < \eps < 1$, we can answer A-Value$(Q, \eps)$ for the discrete \frechet distance in:
\[O \left( \frac{c \cdot m}{\eps} \cdot  \log n \cdot  \left( T_{\eps / 6} + \log \frac{c \cdot m}{\eps}  + \log n \right) \right)  \textnormal{ time}.\]
\end{restatable}

The advantage of our result is that it applies to a variety of metric spaces $\X$, while also improving upon previous \emph{static} algorithms in those spaces. Here, static refers to solutions that do not require preprocessing or building a data structure. For a complete overview of the improvements we make to the state-of-the-art, we refer to Table~\ref{tab:results}.

Our result does not rely upon complicated techniques such parametric search~\cite{gudmundssonfrechet2020, gudmundsson:mapmatching}, higher-dimensional envelopes~\cite{buchin2022efficient}, or advanced path-simplification structures~\cite{driemel2013jaywalking, DriemelPS19, FiltserF21}. Our techniques are not only generally applicable, but also appear implementable (e.g., the authors of~\cite{buchin2022efficient} mention that their result is un-implementable).

\subsection{Corollaries}
\label{section:corollaries}

Theorems~\ref{thm:frechetdecision}+\ref{thm:frechet_value} are the first data structures for computing the \frechet distance between $c$-packed curves. 
Our construction has two novelties: first, we consider $c$-packed curves in \emph{any} metric space $\mathcal{X}$, and only require access to \emph{perceived} distances (distance oracles that can report a $(1+\alpha)$-approximation in $O(T_\alpha)$ time).
Second, we propose a new $1$-TADD technique that can be used to compute the discrete \frechet distance independently of the geodesic ambient space $\mathcal{X}$.
These two novelties imply improvements to previous results, even static results for computing the \frechet distance: 

\subparagraph{Applying perceived distances.}
Our analysis allows us to answer the decision variant for the discrete \frechet distance for any geodesic metric space for which there exist efficient $(1 +\alpha)$-approximate distance oracles (See Oracles~\ref{def:distanceoracles}). 
Combining Theorem~\ref{thm:frechetdecision} with Oracles~\ref{def:distanceoracles} we obtain:

 \begin{restatable}{corollary}{frechetdecisioncor}
 \label{cor:frechetdecision}
 Let $P$ be a $c$-packed curve in a metric space $\X$.
 \begin{itemize}
     \item For $\X = \R^d$ under $L_1, L_2, L_\infty$ metric in the real-RAM model, we can store $P$ using $O(n)$ space and preprocessing, to answer A-decision$(Q, \eps, \rho)$ in $O \left( \frac{c m}{\eps} \cdot \left( d + \log n \right) \right)$ time.
     \begin{itemize}
        \item Improving the static $O( d \cdot \frac{ c n}{ \sqrt{\eps} } + d \cdot c n \log n)$ algorithm by Bringmann and  K{\"u}nnemann~\cite[IJCGA'17]{bringmann2017improved}: making it faster when $n > m$ and making it a data structure.
     \end{itemize}
     \item  For $\X = \R^d$ under $L_2$ in word-RAM, we can store $P$ using $O(n)$ space and $O(n \log n)$ preprocessing, to answer A-decision$(Q, \eps, \rho)$ in $O \left( \frac{c m}{\eps} \cdot \left( d \log \eps^{-1}+ \log n \right) \right)$ worst case time.
     \begin{itemize}
        \item Improving the static expected $O( d^2 \cdot \frac{c n}{ \sqrt{\eps} } + d^2  \cdot c n \log n)$ algorithm by Bringmann and  K{\"u}nnemann~\cite[IJCGA'17]{bringmann2017improved}: saving a factor $d$ and obtaining deterministic guarantees.
     \end{itemize}
     \item For $\X$ an $N$-vertex planar graph under the shortest path metric, we can store $P$ using $O(N^{1+o(1)})$ space and preprocessing, to answer A-decision$(Q, \eps, \rho)$ in $O \left( \frac{c m}{\eps}  \cdot \log^{2+o(1)} N \right)$ time.
    \begin{itemize}
         \item  Generalizing the static  $O \left(N^{1 + o(1)}  + \frac{c \cdot m}{\eps} \log^{2+o(1)}  N \right)$ algorithm by Driemel, van der Hoog and Rotenberg~\cite[SoCG'22]{driemel2023discrete} to a data structure.
    \end{itemize}
     \item For $\X $ an $N$-vertex graph under the shortest path metric, we can fix $\eps$ and store $P$ using $O( \frac{N}{\eps} \log N)$ space and preprocessing, to answer A-decision$(Q, \eps, \rho)$ in $O \left( \frac{c m}{\eps}  \cdot (\eps^{-1} + \log n ) \right)$ time.
     \begin{itemize}
     \item Generalizing the static  $O \left(N^{1 + o(1)}  + \frac{c \cdot m}{\eps} \log^{2+o(1)}  N \right)$ algorithm by Driemel, van der Hoog and Rotenberg~\cite[SoCG'22]{driemel2023discrete} to a data structure.
     \end{itemize}
 \end{itemize}
 \end{restatable}

\subparagraph{Applying the $1$-TADD. }
We can transform the decision variant of the \frechet distance to the optimization variant, by using only a well-separated pair decomposition of $P$ (mapped to $\mathbb{R}^1$) with itself. 
This allows us to answer the optimization variant for the discrete \frechet distance without an exponential dependence on the dimension (speeding up even static algorithms). 
In full generality, combining Theorem~\ref{thm:frechetdecision} with Oracles~\ref{def:distanceoracles} implies:

\begin{restatable}{corollary}{frechetvaluecor}
\label{cor:frechetvalue}
Let $P$ be a $c$-packed curve in a metric space $\X$.
\begin{itemize}
    \item For $\X = \R^d$ in the real-RAM model, we can store $P$ using $O(n)$ space and $O(n \log n)$ preprocessing, to answer A-value$(Q, \eps)$ in $O \left( \frac{c m}{\eps} \log n (d + \log \frac{c \cdot m}{\eps} + \log n \right)$ time.
    \begin{itemize}
        \item Improving the static  
        $O(2^d n + d \cdot \frac{ c n}{ \sqrt{\eps} } + d^2 \cdot c n \log n)$ algorithm by Bringmann and  K{\"u}nnemann~\cite[IJCGA'17]{bringmann2017improved}: removing the exponential dependency on the dimension.
        \item Improving the dynamic $O(  \left( O(\frac{1}{\eps}) \right)^{dm} \log \frac{1}{\eps})$ space solution with $\Tilde{O}(m \cdot d)$ query time by Filtser and Filtser~\cite[SODA'21]{FiltserF21}: allowing $Q$ to have arbitrary length, and using linear as opposed to exponential space and preprocessing time. This applies only when $P$ is $c$-packed.
    \end{itemize}
    \item  For $\X = \R^d$ under the $L_2$ metric in word-RAM, we can store $P$ using $O(n)$ space and $O(n \log n)$ preprocessing, to answer A-value$(Q, \eps)$ in   $O \left( \frac{c m}{\eps} \log n (d \log n + \log \frac{c \cdot m}{\eps} \right)$ time. 
    \begin{itemize}
        \item Improving upon the static expected  $O(2^d n + d^2 \cdot \frac{ c n}{ \sqrt{\eps} } + d^3 \cdot c n \log n)$ algorithm  by Bringmann and  K{\"u}nnemann~\cite[IJCGA'17]{bringmann2017improved}: saving a factor $d 2^d$ with deterministic guarantees.
    \end{itemize}
        \item For $\X$ an $N$-vertex planar graph under the SP metric, we can store $P$ using $O(N^{1+o(1)})$ space and preprocessing, to answer A-value$(Q, \eps)$ in $O \left( \frac{c \cdot m}{\eps} \log n \cdot (\log^{2+o(1)} N + \log \frac{c \cdot m}{\eps}) \right)$ time.
    \begin{itemize}
        \item  Improving the static  $O \left(N^{1 + o(1)} + |E| \log |E| + \frac{c \cdot m}{\eps} \log^{2+o(1)}  N \log |E| \right)$ algorithm by Driemel, van der Hoog and Rotenberg~\cite[SoCG'22]{driemel2023discrete}: making it a data structure.
    \end{itemize}
    \item For $\X $ an $N$-vertex graph under the SP metric, we can fix $\eps$ and store $P$ using $O( \frac{N}{\eps} \log N)$ space and preprocessing, to answer A-value$(Q, \eps)$ in $O \left( \frac{c m}{\eps}  \cdot \log n \cdot (\eps^{-1} + \log \frac{c \cdot m}{\eps} + \log n ) \right)$ time.
    \begin{itemize}
    \item Same improvement as above, except that this result is not adaptive to $\eps$.
    \end{itemize}
    \item For $\X$ an $N$-vertex simple polygon under geodesics, we can store $P$ using $O(N \log N + n)$ space and preprocessing, to answer  A-value$(Q, \eps)$ in $O \left( \frac{c m}{\eps}  \cdot \log n \cdot (\log N + \log \frac{c \cdot m}{\eps} + \log n ) \right)$ time.
\begin{itemize}
    \item No realism-parameter algorithm was known in this setting, because no TADD can be computed in this setting. 
\end{itemize}
    
\end{itemize}
\end{restatable}

We briefly note that all our results are also immediately applicable to subcurve queries: 

\begin{corollary}
All results obtained in Section~\ref{sec:frechet} can answer the subcurve variants of the A-decision and A-value queries for any $i, j \in [n]$ at no additional cost.
\end{corollary}

\section{Simplification and a data structure}
\label{sec:redefine}

To facilitate computations in arbitrary geodesic metric spaces, we modify the definition of $\mu$-simplifications. Our modified definition has the same theoretical guarantees as the previous definition, but works in arbitrary metrics.
Formally, we say that henceforth the $\mu$-simplification is a curve obtained by starting with $p_1$, and recursively adding the first $p_j$ such that the \emph{length} of the subtrajectory $\ell(P[i, j]) > \mu$, where $p_i$ is the last vertex added to the simplified curve. 
This way, our  $\mu$-simplifications (and their computation) are independent of the ambient space and only depend on the edge lengths.

\noindent
We construct a data structure such that for any value $\mu$, we can efficiently obtain $P^\mu$:

\begin{definition}
\label{def:answerdatastructure}
    For any curve $P$ in $\X$ with $n$ vertices, for each $1 < i \leq n$ we create a half-open interval $(\ell(P[1, i-1]), \ell(P[1, i])]$ in $\R^1$. 
This results in an ordered set of $O(n)$ disjoint intervals on which we build a balanced binary tree in $O(n)$ time. 
\end{definition}

 Our new definition and data structure allow us to obtain $P^\mu$ at query time:

\begin{lemma}
\label{lem:rayshooting}
Let $P = (p_1, \ldots, p_n)$ be a curve in $\X$ stored in the data structure of Definition~\ref{def:answerdatastructure}.
For any value $\mu \geq 0$, any pair $(i, j)$ with $i < j$, and any integer $N$ we can report the first $N$ vertices of the discrete $\mu$-simplification $P[i, j]^\mu$ in $O(N \log n)$ time.
\end{lemma}

\begin{proof}
The first vertex of $P[i, j]^\mu$ is $p_i$. We inductively add subsequent vertices. Suppose that we just added $p_x$ to our output. 
We choose the value $a = \ell(P[1, x]) + \mu$.
We binary search in $O(\log n)$ time for the point $p_y$ where the interval  $(\ell(P[1, y-1]), \ell(P[1, y])]$ contains $a$. 
Per definition: the length $\ell(P[x, y]) \geq \mu$. Moreover, for all $z \in (x, y)$ the length $\ell(P[x, z]) < \mu$. 
Thus, $p_y$ is the successor of $p_x$ and we recurse if necessary. 
\end{proof}

\section{The 1-TADD technique }
\label{sec:1TADD}

Let $P$ and $Q$ be curves in $\R^d$ under the Euclidean metric. 
In \cite{DriemelHW12}, they show an algorithm that (given the $\mu$-simplified curves $P^{\mu}$ and $Q^{\mu}$ for $\mu = \eps \rho / 4$) they can decide whether $\FD(P, Q) > \rho$ or $\FD(P, Q) \leq (1 + \eps) \rho$ in $O(d \frac{c(n+m)}{\eps})$ time. 
They then compute a $(1+\eps)$-approximation of $\FD(P, Q)$ through a binary search over TADD($P$, $Q$). 
This approach scales poorly with the dimension $d$ because computing TADD($P$, $Q$) has an exponential dependency on $d$. We alleviate this through our $1$-TADD definiton: 

\begin{definition}
    \label{def:TADD}
    Given $P$, map each vertex $p_i$  to $\lambda_i = \ell(P[1, i])$. Denote by $\Lambda = \{ \lambda_i \}_{i=1}^n$. 
   We define $1$-TADD($P$) as TADD($\Lambda$, $\Lambda$). 
\end{definition}

Our $1$-TADD can be computed in $O(n \log n)$ time using $O(n)$ space~\cite{DriemelHW12}.

\TADD*

\begin{proof}
With slight abuse of notation, we say that A$(P, Q, \eps, \rho)$ is an algorithm that takes as input $P^{\eps \rho /4}$ and $Q^{\eps \rho / 4}$ and outputs either $\FD(P, Q) > \rho$ or $\FD(P, Q) \leq (1 + \eps) \rho$.
We briefly note given our new definition of $\mu$-simplification, \cite{DriemelHW12} present an A$(P, Q, \eps, \rho)$ algorithm with a runtime of $O(d \frac{c(n+m)}{\eps})$ under any $L_p$ metric (as Lemma 4.4 in~\cite{DriemelHW12} immediately works for our simplification definition).  
We use A$(P, Q, \eps, \rho)$ to approximate $\FD(P, Q)$.

 We preprocess $P$ and $Q$ by computing $T_P =$ 1-TADD$(P)$ and $T_Q$ = 1-TADD$(Q)$. 
We denote by $I$ the set of intervals obtained by taking for each $a \in T_P \cup T_Q$ the interval $[4\eps^{-1}a, 8 \eps^{-1}a]$ (we add the interval $[0, 0]$ to $I$).  
Given A$(P, Q, \eps, \rho)$ that runs in $O(d \frac{c(n+m)}{\eps})$ time, we do binary search over $I$. 
Specifically, we iteratively select an interval $[a, b] \in I$ and run A$(P, Q, \eps, \rho)$ for $\rho$ equal to either endpoint. 

Note that for each $\rho$, we may use Lemma~\ref{lem:rayshooting} to obtain both the $\eps \rho / 4$-simplifications of $P$ and $Q$ in $O((n+m)\log (n+m))$ time -- which are required as input for A$(P, Q, \eps, \rho)$. 
We need to do this procedure at most $O(\log n)$ times before we reach one of two cases:
\begin{itemize}
    \item Case 1: There exists $[a, b] \in I$, such that $\FD(P, Q) > a$ and $\FD(P, Q) < (1 + \eps) b$.
We note that by definition of $I$, the values $a$ and $b$ differ by a factor $2$. 
Thus, we may discretize the interval $[a, b]$ into $O(\eps^{-1})$ points that are each at most $\frac{\eps \cdot a}{2}$ apart (note that we implicitly discretize this interval, as an explicit discretization takes $\eps^{-1}$ time).
By performing binary search over this discretized set, we report a $(1 + \eps)$-approximation of $\FD(P, Q)$ by using A$(P, Q, \eps, \rho)$ at most $O(\log \eps^{-1})$ times.
    \item Case 2: There exists no $[x, y] \in I$ such that $\FD(P, Q) > x$ and $\FD(P, Q) < (1 + \eps) y$.
    \begin{itemize}
        \item  Denote by $[a_{max}, b_{max}]$ the right-most interval in $I$. Consider the special case where $\FD(P,Q) > b_{max}$. Since $b_{max} \geq \ell(P), \ell(Q)$ it follows that all $\rho > b_{max}$, $P^{\mu}$ and $Q^{\mu}$ for $\mu = \eps \rho / 4$ are an edge. Computing $\FD(P^{\mu}, Q^{\mu})$ can therefore be done in $O(d)$ time which gives a $(1+\eps)$-approximation of $\FD(P, Q)$.   
    \end{itemize}
    If the special case does not apply then there exist two intervals $[a, b], [e, f] \in I$ such that $\FD(P, Q) > b$ and $\FD(P, Q) \leq (1 + \eps) e$, such that there exists no interval $[x, y] \in I$ that intersects $[b, e]$.   
We claim that  all $\rho_1, \rho_2 \in [b, e]$: $P^{\eps \rho_1 / 4} = P^{\eps \rho_2 / 4}$ and $Q^{\eps \rho_1 / 4} = Q^{\eps \rho_2 /4}$.
\begin{itemize}
    \item Indeed, suppose for the sake of contradiction that $P^{\eps \rho_1 /4} \neq P^{\eps \rho_2 / 4}$. 
Let $\rho_1 < \rho_2$ and choose without loss of generality the smallest $\rho_2$ for which this is the case. Then there must exist a pair $p_i, p_j \in P$ where $\ell(P[i, j]) = \eps \rho_2 / 4$. 
However, the distance $\ell(P[i,j])$ is the distance between $\lambda_i$ and $\lambda_j$ in the curve $\Lambda$ and so there exist $a', b' \in T_P$ with $a' \leq \ell(P[i, j])  \leq b' \leq 2a'$. 
It follows that $\rho_2 = 4\eps^{-1} \ell(P[i, j])$ lies in an interval in $I$ which is a contradiction with the assumption that $\rho_1, \rho_2 \in [b, e]$.  
\end{itemize}

We choose $\rho = e$.
Denote by $X$ the set of reachable cells the Free Space Diagram of $(P^{\eps \rho / 4}, Q^{\eps \rho / 4}, \rho^*)$. 
The set $X$ contains $O( \frac{c(n+m)}{\eps})$ cells~\cite[Lemma 4.4]{DriemelHW12}.
It follows that there are $O(\frac{c(n+m)}{\eps})$ values $\rho'$ for which the reachability of $X$ changes. We compute and sort these to get a sorted set $R$. 

Suppose for some  $\rho' \in [b, \rho]$ that $\FD(P^, Q) \leq \rho'$.
Denote by $F$ an $xy$-monotone path in the Free Space Diagram of $(P^{\eps \rho' /4}, Q^{\eps \rho' / 4}, \rho') = (P^{\eps \rho/4}, Q^{\eps \rho/4}, \rho')$. Per definition, $F$ lies within $X$. 
Thus, we may binary search over the set $R \cap [b, \rho]$ (applying the $\eps$-approximate decider at every step) to compute a $(1 + \eps)$-approximation of $\FD(P,Q)$. \qedhere
        \end{itemize}
 \qedhere
\end{proof}

\section{Approximate distance oracles under the discrete \frechet distance}
\label{sec:approximate}

We want to approximate $\FDd(P, Q)$ for curves $P$ and $Q$ that live in any geodesic ambient space $\X$. 
In most ambient spaces we do not have access to efficient exact distance oracles.
In many ambient spaces however, it is possible to compute for any $\alpha > 0$ some  $(1+\alpha)$-approximate distance oracle. 
This is a data structure $\dist{\alpha}$ that for any two $a,b \in \X$ can report a value $\d(a, b) \in \left[ (1-\alpha) d(a, b), (1 + \alpha) d(a, b) \right]$ in $O(T_\alpha)$ time.
To distinguish between inaccuracy as a result of our algorithm and as a result of our oracle, we refer to $\d(a, b)$ as the \emph{perceived value} (as opposed to an approximate value).  

\begin{oracles}
\label{def:distanceoracles}
We present some examples of approximate distance oracles:
\begin{itemize}
    \item For $\X \subseteq \R^d$ under the $L_1, L_2, L_\infty$ metric in real-RAM we can compute the exact $d(a, b)$ in $O(d)$ time. Thus, for any $\alpha$, we have an oracle $\dist{\alpha}$ with $T_\alpha = O(d)$ query time.
    \item For $\X \subseteq \R^d$ under the $L_2$ metric executed in word-RAM, we can compute $d(a, b)$ in $O(d^2)$ expected time.
    Thus, we have an oracle $\dist{\alpha}$ with $O(d^2)$ expected query time.
    \item For any $\X \subseteq \R^d$ under the $L_2$ metric in word-RAM, we can $(1+\alpha)$-approximate the distance between two points in $T_\alpha = O(d \log \alpha^{-1})$ worst case time using Taylor expansions.
    \item For $\X$ a planar weighted graph, Long and Pettie~\cite{LongPettie} store $\X$ with $N$ vertices using $O\left(N^{1+o(1)}\right)$ space, to answer exact distance queries in  $O\left((\log(N))^{2+o(1)}\right)$ time. 
    \item For $\X$ as an arbitrary weighted graph, Thorup~\cite{thorup2004compact} compute a $(1+\alpha)$-approximate distance oracle in $O(N / \alpha \log N)$ time and space, and with a query-time of $O(1/\alpha)$. 
    \item For $\X$ a simple $N$-vertex polygon, Guibas and Hershberger~\cite{guibas1987optimal} store $\X$ in $O(N \log N)$ time in linear space, and answer exact geodesic distance queries in $O(\left( \log N \right)$ time.
\end{itemize}
\end{oracles}

\noindent
We prove that we may approximately decide the \frechet distance between $P$ and $Q$ using a $(1+\alpha)$-approximate distance oracle (for the discrete \frechet distance). 

\decisionanswer*

\begin{proof}
Per definition of $\dist{\eps / 6}$: $\forall (p, q) \in P \times Q$, $\d(p,q) \in [(1-\frac{1}{6}\eps) d (p,q),  (1+\frac{1}{6}\eps) d(p,q)]$.

It follows from $0 < \eps < 1$ that:
\[
\forall (p, q) \in P \times Q : \quad d (p,q) \leq \left(1 + \frac{1.1}{6}\eps \right) \d(p, q) \quad \wedge \quad \d(p,q) \leq \left(1+\frac{1}{6}\eps \right) d(p,q).
\]

\textbf{Suppose that $\FDd^\circ(P^\mu, Q) \leq \rho^*$.} There exists a (monotone) discrete walk $F$ through $P^\mu \times Q$ such that for each $(i, j) \in F$: 
$\d( P^\mu[i], q_j) \leq \rho^* = (1 + \frac 1 2 \eps)\rho$. 
It follows that:
\[
d(P^\mu[i],q_j) \leq \left(1+\frac{1.1}{6} \eps \right) \d (P^\mu[i],q_j) \leq \left(1+\frac{1.1}{6} \eps \right) \left(1 + \frac 1 2 \eps \right)\rho \leq \left(1+ \frac 5 6 \varepsilon \right) \rho. \]

We will prove that this implies $\FDd(P, Q) \leq (1 + \eps)\rho$. We use $F$ to construct a discrete walk $F'$ through $P \times Q$. 
For each consecutive pair $(a, b), (c, d) \in F$ note that since $F$ is a discrete walk, $P^\mu[a]$ and $P^\mu[c]$ are either the same vertex or incident vertices on $P^\mu$. 
Denote by $P_{ac}$ the vertices of $P$ in between $P^\mu[a]$ and $P^\mu[c]$.
It follows that:
\[
 \forall p' \in P_{ac}: \quad  
d(p',q_b) \leq d \left( P^\mu[a], q_b \right) + \mu \leq (1+ \frac 5 6 \varepsilon) \rho + \frac 1 6 \varepsilon \rho = (1+\varepsilon) \rho.
\]

\noindent
Now consider the following sequence of pairs of points: 

$
L_{ac} = (P^\mu[a], q_b) \cup \{ (p', q_b) \mid p' \in P_{ac}  \} \cup (P^\mu[c], q_d).
$
We add the lattice points corresponding to $L_{ac}$ to $F'$. It follows that we create a discrete walk $F'$ in the lattice $|P| \times |Q|$ where for each $(i, j) \in F'$: $d(p_i, q_j) \leq (1+\eps)\rho$. Thus, $\FDd(P, Q) \leq (1 + \eps)\rho$.

\textbf{Suppose otherwise that $\FDd(P, Q) \leq \rho$.}
We will prove that $\FDd^\circ(P^\mu, Q) \leq \rho^*$. 
Indeed, consider a discrete walk $F'$ in the lattice $|P| \times |Q|$ where for each $(i, j) \in F'$: $d(p_i, q_j) \leq \rho$. 
We construct a discrete walk $F$ in $|P^\mu| \times |Q|$. Consider each $(i, j) \in F'$,
If $p_i = P^\mu[a]$ for some integer $a$, we add $(a, j)$ to $F$.
Otherwise, denote by $P^\mu[a]$ the last vertex on $P^\mu$ that precedes $p_i$:
we add $(a, j)$ to $F$. 
Note that per definition of $\mu$-simplification, $d(P^\mu[a], q_j) \leq d(p_i, q_j) + \mu \leq (1 + \frac 1 6 \eps)\rho$.
It follows from the definition of our approximate distance oracle that $\d(P^\mu[a], q_j) \leq (1 + \frac 1 6 \eps)(1 + \frac 1 6 \eps) \rho < (1 + \frac 1 2 \eps)\rho = \rho^*$.
Thus, we may conclude that $\FDd^\circ(P^\mu, Q) \leq \rho^*$. 
\end{proof}

\section{Approximate Discrete \frechet distance}
\label{sec:frechet}

We denote by $\dist{\alpha}$ a $(1+\alpha)$-approximate distance oracle over the geodesic metric space $\X$.
Our input is some curve $P = (p_1, \ldots, p_n)$ in $\X$ which is $c$-packed in $\X$.
We preprocess $P$ to:
\begin{itemize}
    \item answer A-decision$(Q, \eps, \rho)$ for any curve $Q = (q_1, \ldots, q_m)$, $\rho > 0$ and $0 < \eps < 1$, 
    \item answer A-value$(Q, \eps)$ for any curve $Q = (q_1, \ldots, q_m)$ and $0 < \eps < 1$.
\end{itemize}

We obtain this result in four steps.
In Section~\ref{sec:approximate}, we showed that we can answer A-decision$(Q, \eps, \rho)$ through comparing if the perceived \frechet distance $\FDd^\circ(P^\mu, Q) \leq \rho^*$ for conveniently chosen $\mu$ and $\rho^*$. 
In Section~\ref{sub:matrix} we define what we call the perceived free-space matrix. 
This is a $(0, 1)$-matrix $M^{A \times Q}_{\rho^*}$ for any two curves $A$ and $Q$ and any $\rho^* > 0$. We show that if $A$ is the $\mu$-simplified curve $P^\mu$ for some convenient $\mu$, then the number of zeroes in $M^{P^\mu \times Q}_{\rho^*}$ is bounded.

In Section~\ref{sub:decision}, we show a data structure that stores $P$ to answer A-decision$(Q, \eps, \rho)$. 
We show how to cleverly navigate $M^{P^\mu \times Q}_{\rho^*}$ for conveniently chosen $\mu$ and $\rho^*$. The key insight in this new technique, is that we may steadily increase $\rho^*$ whilst navigating the matrix.
Finally, we extend this solution to answer A-value$(Q, \eps)$.

\subsection{Perceived free space matrix and free space complexity}
\label{sub:matrix}

We define the perceived free space matrix to help answer A-decision queries. 
Given two curves $(A, Q)$ and some $\rho$, we construct an $|A| \times |Q|$ matrix which we call the perceived free space matrix $M^{A \times Q}_\rho$. The $i$'th column corresponds to the $i$'th element $A[i]$ in $A$. 
We assign to each matrix cell $M^{P \times Q}_\rho[i, j]$ the integer $0$ if $\d(A[i], q_j) \leq \rho$ and integer $1$ otherwise.

\begin{observation}
\label{obs:zeroes}
For all $\rho^* \geq 0$, and curves $A$ and $Q$, the perceived discrete \frechet distance $\FDd^\circ(A, Q)$ between $A$ and $Q$ is at most $\rho^*$ if and only if there exists an ($xy$-monotone) discrete walk $F$ from $(1, 1)$ to $(|A|, |Q|)$ where $\forall (i, j) \in F$, $M^{A \times Q}_{\rho^*}[i, j] = 0$.
\end{observation}

\subparagraph{Computing $\FDd^\circ(P^\mu, Q)$.}
Previous results
upper bound, for any choice of $\rho$, the number of zeroes in the FSM between $P^{\eps \rho}$ and $Q^{\eps \rho}$. 
We instead consider the perceived FSM, and introduce a new parameter $k \geq 1$ to enable approximate distance oracles. 
For any value $\rho$ and some simplification value $\mu \geq \frac{\eps \rho}{k}$, we upper bound the number of zeroes in the perceived FSM $M^{P^{\mu} \times Q}_{\rho^*}$ for the conveniently chosen $\rho^* = (1 + \frac{\eps}{2})\rho$:

\begin{lemma}
\label{lem:zeroes_per_row}
Let $P = (p_1, \ldots, p_n)$ be a $c$-packed discrete curve in $\X$.
For any $\rho > 0$ and $0 < \eps < 1$, denote $\rho^* = (1 + \frac{\eps}{2})\rho$.
For any $k \geq 1$, denote by $P^\mu$ its $\mu$-simplified curve for $\mu \geq \frac{\eps \rho}{k}$. 
For any curve $Q = (q_1, \ldots, q_m) \subset \X$ the matrix $M^{P^{\mu} \times Q}_{\rho^*}$ contains at most
$8 \cdot \frac{c \cdot k}{\eps} $ zeroes per row.
\end{lemma}

\begin{figure}[t]
  \centering
  \includegraphics[]{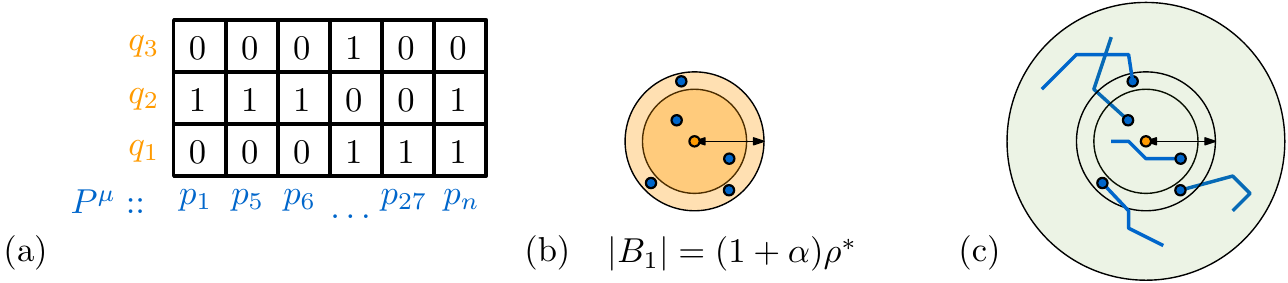}
  \caption{
(a) For some value of $\mu$, the $\mu$-simplified curve is $(p_1, p_5, p_6, \ldots, p_{27}, p_n)$. 
We show the matrix $M^{P^\mu \times Q}_{\rho^*}$.
(b) For the point $q_3$, we claim that there are more than $Z = 8\frac{ck}{\eps} $ zeroes in its corresponding row. Thus, the ball $B_1$ with radius $(1+\alpha)\rho^*$ contains more than $Z$ points.
(c) For each of these points, there is a unique segment along $P$ contained in the ball $B_2$.
  }
  \label{fig:balls}
\end{figure}

\begin{proof}
The proof is by contradiction.
Suppose that the $j$'th row of $M^{P^{\mu} \times Q}_{\rho^*}$ contains strictly more than $8 \cdot  \frac{c \cdot k}{\eps}$ zeroes. 
Let $P_0 \subset P^\mu$ be the vertices corresponding to these zeroes.
Consider the ball $B_1$ centered at $q_j$ with radius $|B_1| = (1 + \alpha)\rho^*$ and the ball $B_2$ with radius $2 |B_1|$ (Figure~\ref{fig:balls}). 
Each $p_i \in P_0$ must be contained in $B_1$ and thus $d(p_i, q_j) \leq (1 + \alpha)\rho^*$.
For each $p_i \in P_0$ denote by $S_i$ the contiguous sequence of vertices of $P^\mu$ starting at $p_i$ of length $\mu$. Observe that since $\eps < 1$: $S_i \subset B_2$.
Per definition of simplification, each $S_i$ are non-coinciding subcurves.
This lower bounds $\ell(P \cap B_2)$:
\[
\ell( P \cap B_2) \geq \sum_{p \in P_0} \ell(S_i) =  \sum_{p \in P_0} \mu > 2(1 + \alpha) (1 + \frac{\eps}{2}) \cdot \frac{c \cdot k}{\eps} \cdot \mu \geq 2 (1 + \alpha) \cdot c \cdot \rho^* \geq c \cdot |B_2|,
\]
where $2(1 + \alpha) (1 + \frac{\eps}{2}) \cdot \frac{c \cdot k}{\eps} < 8 \cdot \frac{c \cdot k}{\eps}$ (since $\alpha < 1$ and $\varepsilon < 1$) -- contradicting $c$-packedness. 
\end{proof}


\subsection{A data structure for answering \texorpdfstring{A-decision$(Q, \eps, \rho)$}{A-decision(Q, e, r)}}
\label{sub:decision}

We showed in Sections~\ref{sec:approximate} and \ref{sub:matrix} that for any $c$-packed curve $P$, $\rho > 0$ and $0 < \eps < 1$ we can choose suitable values $\frac{\eps \rho}{k} \leq \mu \leq \frac{\eps \rho}{6}$ to upper bound the number of zeroes in $M^{P^\mu \times Q}_{\rho^*}$. Moreover, for $\rho^* = (1 + \frac{1}{\eps})\rho$ we know that comparing $\FDd^\circ(P^\mu, Q) \leq \rho^*$ implies an answer to A-decision$(Q, \eps, \rho)$.

We now define a data structure, so that for any $\mu$ and any $(i, j)$ we can report the $\mu$-simplification of $P[i, j]$ in $O(|P[i, j]|)$ time. We use this to answer the decision variant.

\frechetdecision*

\begin{proof}
We store $P$ in the data structure of Definition~\ref{def:answerdatastructure} using $O(n)$ space and preprocessing time. 
Given a query A-decision$(Q, \eps, \rho)$ we choose $\alpha = \frac{1}{6} \eps$, $\rho^* = (1 + \frac{1}{2} \eps)$ and $\mu = \frac{\eps \rho}{6}$. 
We test if $\FD^\circ(P^\mu, Q) \leq \rho^*$. By Lemma~\ref{lem:decision_answer}, if $\FDd^\circ(P^\mu, Q) \leq \rho^*$ then $\FDd(P, Q) \leq (1 + \eps) \rho$ and otherwise $\FDd(P, Q) > \rho$. 
We consider the matrix $M^{P^\mu \times Q}_{\rho^*}$.

By Observation~\ref{obs:zeroes}, $\FDd^\circ(P^\mu, Q) \leq \rho^*$ if and only if there exists a discrete walk $F$ from $(1, 1)$ to $(|P^\mu|, |Q|)$ where for each $(i, j) \in F$: $M^{P^\mu \times Q}_{\rho^*}[i, j] = 0$.
We will traverse this matrix in a depth-first manner as follows: 
starting from the cell $(1, 1)$, we test if $M^{P^\mu \times Q}_{\rho^*}[1, 1] = 0$. 
If so, we push $(1, 1)$ onto a stack.
Each time we pop a tuple $(i, j)$ from the stack, we inspect their $O(1)$ neighbors $\{ (i + 1, j), (i, j+1), (i+1, j+1) \}$. 
If $M^{P^\mu \times Q}_{\rho^*}[i', j'] = 0$, we push $(i', j')$ onto our stack.
It takes $O(\log n)$ time to obtain the $i+1$'th vertex of $P^\mu$, and $O(T_{\eps / 6})$ to determine the value of e.g., $M^{P^\mu \times Q}_{\rho^*}[i+1, j]$.
Thus each time we pop the stack, we spend $O((T_{\eps / 6}) + \log n)$ time.

By Lemma~\ref{lem:zeroes_per_row} (noting $\eps < 1$ and setting $k = 6$), we push at most $O(\frac{c m}{\eps})$ tuples onto our stack. Therefore, we spend $O( \frac{c m}{\eps}(T_{\eps / 6} + \log n))$ total time. 
By Observation~\ref{obs:zeroes}, $\FDd^\circ(P^\mu, Q) \leq \rho^*$ if and only if we push $(|P^\mu|, |Q|)$ onto our stack. 
We test this in $O(1)$ additional time per operation. Thus, the theorem follows.
\end{proof}

\subsection{A data structure for answering \texorpdfstring{A-value$(Q, \eps)$}{A-value(Q, e)}}
\label{sub:value}
Finally, we show how to answer the A-value$(Q, \eps, \rho)$ query. 
At this point, we could immediately apply Theorem~\ref{thm:TADD} to answer A-value$(Q, \eps)$ at the cost of a factor $O(\log n + \log \eps^{-1})$. However, for the discrete \frechet distance we show that the factor $O(\log \eps^{-1})$ can be avoided. To this end, we leverage the variable $k \geq 1$ introduced in the definition of $\mu \geq \frac{\eps \rho}{k}$: 

\frechetvalue*

\begin{proof}
We preprocess $P$ using Lemma~\ref{lem:rayshooting} in $O(n)$ space and time. 
We store $P$ in the data structure of Definition~\ref{def:TADD}. 
This way, we obtain $T$ = TADD($\Lambda$, $\Lambda$) where $\Lambda$ is the curve $P$ mapped to $\mathbb{R}^1$.
We denote for all $s \in T$  by $I_s = [ c_s, 2 \cdot c_s ]$ the corresponding interval and obtain a sorted set of intervals $\mathcal{I} = \{ I_s \}$.

Given a query $(Q, \eps)$,  we set $\alpha \gets \eps / 6$ and obtain $\dist{\alpha}$. 
We (implicitly) rescale each interval $I_i \in \mathcal{I}$ by a factor $\frac{6}{\eps}$, creating for $I_s$ the interval $I_s^\eps = [ \frac{6 \cdot c_s}{\eps}, \frac{12 \cdot c_s}{\eps}]$. 
This creates a sorted set $\mathcal{I}^\eps$ of pairwise disjoint intervals. Intuitively, these are the intervals over $\R^1$ where for $\rho \in I_s^\eps$, the $\mu$-simplification $P^\mu$ for $\mu = \frac{\eps \rho}{6}$ may change. 

We binary search over $\mathcal{I}^\eps$. For each boundary point $\lambda$ of an interval $I_s^\eps$ we query A-decision$(Q, \eps, \lambda)$: discarding half of the remaining intervals in $\mathcal{I}^\eps$.
It follows that in $O( \frac{c \cdot m}{\eps} \cdot \log n \cdot (T_{\eps / 6} + \log n))$ time, we obtain one of two things:
\begin{enumerate}[a)]
    \item an interval $I_s^\eps$ where $\exists \rho^* \in I_s^\eps$ that is a $(1 + \eps)$-approximation of $\FDd(P, Q)$, or
    \item a maximal interval $I^*$ disjoint of the intervals in $\mathcal{I}^\eps$ where $\exists \rho^* \in I^*$ that is a $(1 + \eps)$-approximation of $\FDd(P, Q)$. 
\end{enumerate}

\noindent
Denote by $\lambda$ the left boundary of $I_s^\eps$ or $I^*$: it lower bounds $\FDd(P, Q)$.
Note that if $I^*$ precedes all of $\mathcal{I}^\eps$, $\lambda = 0$.
We now compute a $(1 + \eps)$-approximation of $\FDd(P, Q)$ as follows:


\subparagraph*{FindApproximation$(\lambda)$:}
\begin{enumerate}
\item Compute  $C = \frac{ \d(p_1, q_1)}{(1 + \frac{1}{2} \eps)}$.
    \item Initialize $\rho^* \gets (1 + \frac{1}{2} \eps) \cdot \max \{ C, \lambda \}$ and set a constant $\mu \gets \frac{\eps}{6} \cdot \lambda$.
    \item Push the lattice point $(1, 1)$ onto a stack. 
    \item Whilst the stack is not empty do:
    \begin{itemize}
        \item Pop a point $(i, j)$ and consider the $O(1)$ neighbors $(p_a, q_b)$ of $(p_i, q_j)$ in $M^{P^\mu \times Q}_{\rho^*}$:
       \begin{itemize}
           \item If $\d(p_a, q_b) \leq \rho^*$,  push $(a, b)$ onto the stack.
           \item Else, store $\d(p_a, q_b)$ in a min-heap.
       \end{itemize} 
       \item If we push $(p_n, q_m)$ onto the stack do:
       \begin{itemize}
           \item Output $\nu = \frac{\rho^*}{(1 + \frac{1}{2}\eps)}$.
       \end{itemize}
    \end{itemize}
    \item If the stack is empty, we extract the minimal $\d(p_a, p_b)$ from the min-heap.
    \begin{itemize}
        \item Update $\rho^* \gets (1 + \frac{1}{2}\eps) \cdot \d(p_a, q_b)$, push $(a, b)$ onto the stack and go to line 4.
    \end{itemize}
\end{enumerate}

\subparagraph*{Correctness.}
Suppose that our algorithm pushes $(p_n, q_m)$ onto the stack and let at this time of the algorithm, $\rho^* = (1 + \frac{1}{2} \eps) \nu$. 
Per definition of the algorithm, $\nu \geq \lambda$ is the minimal value for which the matrix $M^{P^\mu \times Q}_{\rho^*}$ contains a walk $F$ from $(1, 1)$ to $(n, m)$ where for each $(i, j) \in F$: 
$M^{P^\mu \times Q}_{\rho^*}[i, j] = 0$. 
Indeed, each time we increment $\rho^*$ by the minimal value required to extend any walk in $M^{P^\mu \times Q}_{\rho^*}$. Moreover, we fixed $\mu \gets \frac{\eps}{6} \lambda$ and thus $\mu \leq \frac{\eps}{6} \nu$.
Thus we may apply Lemma~\ref{lem:decision_answer} to defer that $\nu$ is the minimal value for which $\FDd(P, Q) \leq (1 + \eps) \nu$.

\subparagraph*{Running time.}
We established that the binary search over $\mathcal{I}^\eps$ took $O( \frac{c \cdot m}{\eps} \cdot \log n \cdot (T_{\eps / 6} + \log n))$ time. We  upper bound the running time of our final routine. 
For each pair $(p_i, q_j)$ that we push onto the stack we spend at most $O(T_{\eps / 6} + \log \frac{c \cdot m}{\eps} + \log n)$ time as we: 
\begin{itemize}
    \item Obtain the $O(1)$ neighbors of $(p_i, q_j)$ through our data structure in  $O(\log n)$ time,
    \item Perform $O(1)$ distance oracle queries in $O(T_{\eps / 6})$ time, and
    \item Possibly insert $O(1)$ neighbors into a min-heap. The min-heap has size at most $K$: the number of elements we push onto the stack. Thus, this takes $O(\log K)$ insertion time.
\end{itemize}
What remains is to upper bound the number of items we push onto the stack. 
Note that we only push an element onto the stack, if for the current value $\rho^*$ the matrix $M^{P^\mu \times Q}_{\rho^*}$ contains a zero in the corresponding cell. We now refer to our earlier case distinction.

Case (a): 
Since $\eps < 1$ we know that $\rho^* \in [\lambda, 4 \cdot \lambda]$.
We set $\mu = \frac{\eps}{6} \lambda$. So $\mu \geq \frac{1}{k}  \eps  \rho^*$ for $k = 24$. 
Thus, we may immediately apply Lemma~\ref{lem:zeroes_per_row} to conclude that we push at most $O(\frac{c \cdot m}{\eps})$ elements onto the stack. 

Case (b):
Denote  by $\gamma = \frac{\eps}{6} \nu$.
Per definition of our re-scaled intervals, the open interval $(\mu, \gamma)$ does not intersect with any interval in the non-scaled set $\mathcal{I}$. 
It follows that $P^\mu = P^\gamma$ and that for two consecutive vertices $p_i, p_l \in P^\mu$: $\ell(P[i, l]) > \gamma$.
From here, we essentially redo Lemma~\ref{lem:zeroes_per_row} for this highly specialized setting.
The proof is by contradiction, where we assume that for $\rho^* = (1 + \frac{\eps}{2}) \nu$ there are more than $8 \cdot 6 \cdot \frac{c}{\eps}$ zeroes in the $j$'th row of $M^{P^\mu \times Q}_{\rho^*}$. 
Denote by $P_0 \subset P^\mu$ the vertices corresponding to these zeroes. 
We construct a ball $B_1$ centered at $q_j$ with radius $2 \rho^*$ and a ball $B_2$ with radius $2 |B_1|$. 
We construct a subcurve $S_i$ of $P$ starting at $p_i \in P_0$ of length $\gamma$. 
The critical observation is, that our above analysis implies that all the subcurves $S_i$ do not coincide (since each of them start with a vertex in $P^\mu$). 
Since $\eps < 1$, each segment $S_i$ is contained in $B_2$.
However, this implies that $B_2$ is not $c$-packed since: $\ell(P \cap B_2) \geq \sum_{i} \ell(S_i) = \sum_i \gamma > 8 \cdot 6 \frac{c}{\eps} \gamma \geq  4 \cdot c \cdot \rho^* \geq 2 \cdot c \cdot |B_2|$.
Thus, we always push at most $O(\frac{c \cdot m}{\eps})$ elements onto our stack and this implies our running time.
\end{proof}

\section{Extension/truncation updates, and subcurve queries}

\label{sec:dynamic}

We expand our techniques to include a limited type of updates to the curve $P$.
Our data structure in Section~\ref{sec:frechet} can be constructed, stored and queried with deterministic worst-case time and space. 
Next, we augment this structure to allow for updates where we may append a new vertex to $P$ (appending it to either the start or end of $P$), or reduce $P$ by one vertex (removing either the start or end vertex).
We assume that after each update, the curve remains $c$-packed (i.e., we are given some value $c$ beforehand which may over-estimate the $c$-packedness of the curve, but never underestimate it). 
Alternatively, the query may include the updated value for $c$, in which case our data structure is adaptive to $c$. 
For simplicity, we assume that the update supplies us with the exact distance between the vertices affected by the update. 
In full generality, our algorithm may query the (approximate) distance oracles for these distance values and work with perceived lengths.
Our data structure requires three components: a $\mu$-simplification data structure, a TADD, and an estimate $c'$ of the $c$-packedness. 
By assumption, we have access to the $c$-packedness estimate.
Therefore, it suffices to update the $\mu$-simplification and the TADD:

\subparagraph{The $\mu$-simplification data structure. }
Our $\mu$-simplification data structure stores for every $i$, the half-open interval $(\ell(P[1, i-1]), \ell(P[1, i]) ]$ in a balanced binary tree.
The key observation here is that our algorithm does not require us to store each interval explicitly. 
Instead, it suffices to compute the width of each interval in the leaves of the tree. 
In each interior node $\nu$ of the tree, we store the sum of the width of all the leaves in the subtree rooted at $\nu$.

\begin{lemma}[Analogue to Lemma~\ref{lem:rayshooting}]
\label{lemma:dyna}
We can update our $\mu$-simplification data structure in $O(\log n)$ worst case time after \emph{any} insertion and deletion in $P$. Afterwards, we can report the first $N$ vertices on the $\mu$-simplification $P^\mu$ in $O(N \log n)$ time.
\end{lemma}

\begin{proof}
    After an update, we identify the leaves affected by this update in $O(\log n)$ time. We compute the new leaves of our tree in $O(1)$ time. 
    Having inserted them, we rebalance the tree.
    For any given value $\mu$, we now obtain the first $N$ vertices of $P^\mu$ as follows:  first, we add the vertex $p_1$. Now we apply induction, where we assume that we have just inserted $p_i$, and have a pointer to the leaf corresponding to $p_i$.
From the leaf of $p_i$, we walk upwards until we are about to reach a node $\nu$ where the value stored at $\nu$ exceeds $\mu$. From there, we walk downwards in symmetric fashion and identify the vertex $p_j$ succeeding $p_i$. 
\end{proof}

\subparagraph{The TADD.}
 Fischer and Har-Peled~\cite{fischer2005dynamic} study the following problem: 
 given a point set $M$ in $\R^d$ and a constant $\eps$, maintain a WSPD of $M \times M$. For ease of exposition, we cite their more general result for our case where we have a point set in $\R^1$:

\begin{lemma}[Theorem~12+13 in \cite{fischer2005dynamic}]
\label{lem:TADD}
    Let $L$ be a list of Well-separated pairs of $M \times M$ for an $n$-point set $M$ in $\R^1$. With high probability, in time $O(\log^2 n)$, one can modify $L$ to yield $\hat{L}$: the list of well-separated pairs of $M \times M$ after an insertion or deletion in $M$.
\end{lemma}
 
We maintain a WSPD of $P' \times P'$: where $P'$ is every point $p_i \in P$ mapped to $\ell(P[1, i])$. 
Thus, inserting a point into $P$, may change $O(n)$ points in $P'$. 
However, note that pairwise distances of $P'$ are preserved under translation of $P'$. 
We maintain a WSPD of $M \times M$ where $M$ originally is $P'$. 
Suppose that we receive a new point $p_0$ at the head of $P$, which is at some distance $e$ from $p_1$. 
We insert into $M$ a new value, which is the value corresponding to $p_1$, minus $e$, and update the WSPD in $O(\log^2 n)$ time with high probability. 
Reducing $P$ or appending/reducing $P$ at its tail can be accomplished in the same way.
Using the 1:1 correspondence between a 1-dimensional WSPD and a TADD, we maintain a TADD of $M$ which is a translated version of $P'$.
Thus, Lemma~\ref{lemma:dyna} and~\ref{lem:TADD} imply Theorem~\ref{thm:dynamic}.

\begin{restatable}{theorem}{dynamic}
\label{thm:dynamic}
We support appending to or reducing $P$ in $O(\log^2 n)$ time (w.h.p).     
\end{restatable}

\section{Map matching}
\label{app:mapmatching}

The technique in Section~\ref{sub:value} avoids the use of parametric search when minimizing the Fr\'echet distance. We apply this technique to map matching under the discrete Fr\'echet distance \emph{and} the discrete Hausdorff distance.

In the map matching problem, the input is a Euclidean graph (a graph $P$ with $n$ vertices embedded in $\R^2$).
The goal is to preprocess $P$, so that any query curve $Q$ in $\R^2$ can be `mapped' onto $P$.
That is, we want to find a path $\pi$ in the graph $P$, such that the distance $\FDd(\pi, Q)$ is minimized.

Under the continuous Fr\'echet distance, it was previously was shown that one can preprocess a $c$-packed graph in quadratic time and nearly-linear space for efficient approximate map matching queries~\cite{gudmundsson:mapmatching}. The query time is $O(m \log m \cdot (\log^4 n + c^4 \eps^{-8} \log^2 n))$, where $n$ and $m$ are the number of edges in the graph and query curve, respectively. 

We consider changing the distance metric to the discrete \frechet distance and the discrete Hausdorff distance.
Using our techniques,  the query time in these settings improves to $O(m (\log n + \log \eps^{-1}) \cdot (\log^2 n + c^2 \varepsilon^{-4} \log^1 n))$, where $n$ and $m$ are the number of edges in the graph and query curve, respectively. In particular, we reduce the polynomial dependence on $c$, $\log n$ and $\log m$ in the query time.
We obtain this, by replacing the parametric search in~\cite{gudmundsson:mapmatching} by our TADD-technique. 
Since $P$ is a graph and not a single curve, we do something slightly different as we compute a two-dimensional TADD on $P$ with itself (in the exact same manner as in Appendix~\ref{app:hausdorff}). 

\begin{restatable}{theorem}{mapmatching}
\label{thm:mapmatching}
Let $\X = \mathbb R^2$ and $P$ be a $c$-packed graph in $\X$. We can store $P$ using $O(n \log^2 n + c \varepsilon^{-4} n \log n \cdot \log \eps^{-1})$ space and $O(c^2 \varepsilon^{-4} n^2 \log^2 n \cdot \log \eps^{-1})$ preprocessing time, such that for any curve $Q$ in $\X$, we can return, in $O(m (\log n + \log \eps^{-1}) \cdot (\log^2 n + c^2 \varepsilon^{-4} \log^1 n))$ time a $(1+\varepsilon)$-approximation of $\min_{\pi \in P} \FDd(\pi,Q)$ (or $\min_{\pi \in P} \HD(\pi,Q)$).
\end{restatable}

\begin{proof}[Proof (Sketch)]
We explain how to modify the proofs by Gudmundsson et al.~\cite{gudmundsson:mapmatching} to use our techniques instead.
We build the data structure in the same way as in Gudmundsson et al.~\cite{gudmundsson:mapmatching}. We modify Lemma~13 of~\cite{gudmundsson:mapmatching} so that, instead of computing the the continuous Fr\'echet distance in $O(n \log n)$ time using a free space diagram, we compute the discrete Fr\'echet distance in $O(n)$ time. 
Specifically, in Lemma~13, the authors show how to compute a map matching between $P$ and $\overline{ab}$ between  any two vertices $a, b \in P$.
We note that the dynamic program to achieve this has linear running time instead of the traditional quadratic running time, since one of the two curves has only two vertices. 

Therefore, Theorem~3 of~\cite{gudmundsson:mapmatching} implies that one can construct a data structure using $O(n \log^2 n + c \varepsilon^{-4} n \log n \cdot \log \eps^{-1})$ space and $O(c^2 \varepsilon^{-4}  n^2 \log^2 n \cdot \log \eps^{-1})$ preprocessing time, to answer the \emph{decision variant} of the query problem in $O(m(\log^2 n + c^2 \varepsilon^{-4} \log n))$ time. This holds for both the discrete Fr\'echet distance and the discrete Hausdorff distance. 

What remains, is to apply the decision variant to efficiently obtain a $(1 + \eps)$-approximation. 
Previously in~\cite{gudmundsson:mapmatching}, parametric search was applied, which introduces a factor of $O(\log m)$ and squares the polynomial dependencies on $\log n$, $c$ and $\varepsilon^{-1}$. We avoid parametric search.

In preprocessing time, we precompute a Voronoi diagram on $P$ in $O(n \log n)$ time, and a 2-dimensional TADD on $P$ and $P$ in $O(n \log n)$ time~\cite{har2011geometric}. The TADD partitions $P \times P$ into $O(n)$ sets $(P_s, P_s')$ where for each $s$ there exists a distance $c_s$ such that for all pairs of points $(p_i, p_j)$ contained in  $(P_s, P_s')$, we have  $c_s \leq d(p_i, p_j) \leq 2c_s$. We sort the values $c_s$ and define $H_s$ to be the interval $[\frac 1 2 \cdot c_s, 4 \cdot c_s]$. This gives a sorted set $\mathcal{H}$.

Given a query $(Q, \eps)$ we first do the following: 
we compute the Hausdorff distance from $Q$ to $P$ in $O(m \cdot \log n)$ time (by querying for every point in $Q$, the Voronoi diagram in $P$ and taking the maximum).
Let this Hausdorff distance be $\lambda$, we create an interval $H^* = [\lambda, 2 \lambda]$.

Let $\rho$ be the map matching distance between $Q$ and $P$. We claim that $\rho$ contained in either an interval in $\mathcal{H}$ or in $H^*$.
Indeed, consider for each $q \in Q$ a disk with radius $\rho$ centered at $q_j$.
Each of these disks must contain at least one point of $P$ (else, we cannot map $q_j$ to $P$ with distance at most $\rho$).
Moreover, there must exist at least one $q_j \in Q$ which has a point $p_i$ on its border (else, we may decrease $\rho$ and still maintain a discrete map matching to $P$). 
Consider the disk $D_j$ centered at $q_j$ with radius $\frac{1}{2} \rho$. There exist two cases:
\begin{enumerate}[a)]
    \item $D_j$ contains a point $p_l \in P$. 
    Then $\rho \in [\frac{1}{2} \cdot d(p_l, p_i), 4 \cdot d(p_l, p_i)]$ and so $\rho$ is in $\mathcal{H}$.
    \item $D_j$ contains no points in $P$. 
    Then $\HD(Q \to P)$ is at least $\frac{1}{2} \rho$. 
    Observe that $\HD(Q \to P)$ is upper bounded by $\rho$, and so $\rho \in H^*$. 
\end{enumerate}

We now use decision variant to binary search over $\mathcal{H}$ to find an interval containing $\rho$ (we check $H^*$ separately).
It follows that we obtain an interval $[\alpha, \beta]$ with $\rho \in [\alpha, \beta]$ and $\frac{\beta}{\alpha} \in O(1)$. We use the standard approach to refine the interval into a $(1+\varepsilon)$-approximation using $O(\log \eps^{-1})$ decision queries. This completes the description of the query procedure. 

Finally, we perform an analysis of the preprocessing time and space, and the query time. The preprocessing is dominated by the construction of Theorem~3 of~\cite{gudmundsson:mapmatching}, which requires $O(n \log^2 n + c \varepsilon^{-4}\log(1/\varepsilon) n \log n)$ space and $O(c^2 \varepsilon^{-4} \log(1/\varepsilon) n^2 \log^2 n)$ preprocessing time. The query procedure consists of a binary search, which requires $O(\log n)$ applications of the query decider to identify the interval $I_s$ or $I_j$. We require a further $O(\log \eps^{-1})$ applications of the query decider to refine the $O(1)$-approximation to a $(1+\varepsilon)$-approximation. Since the decider takes $O(m(\log^2 n + c^2 \varepsilon^{-4} \log n))$ time per application, the overall query procedure can be answered in $O(m \cdot (\log n + \log \eps^{-1}) (\log^2 n + c^2 \varepsilon^{-4} \log n))$ time. 
\end{proof}

\begin{corollary}
\label{cor:mapmatching}
   For the discrete \frechet distance, we improve the query time from~\cite[SODA'23]{gudmundsson:mapmatching} from 
 $O(m \log m \cdot (\log^4 n + c^4 \eps^{-8} \log^2 n))$ to
$O(m  (\log n + \log \eps^{-1}) \cdot (\log^2 n + c^2 \varepsilon^{-4} \log n))$.
\end{corollary}

\section{Hausdorff}
\label{app:hausdorff}

The Hausdorff distance can be defined between any two sets $P$ and $Q$. The discrete directed Hausdorff distance from $P$ to $Q$ is:
\[
\HD(P \tor Q) := \max_{p_i \in P} \min_{q_j \in Q} d(p_i, q_j).
\]
The discrete Hausdorff distance $\HD(P, Q)$ is the maximum of $\HD(P \tor Q)$ and $\HD(Q \tor P)$.

This section is dedicated to showing that our approach also works to compute a $(1 + \eps)$-approximation of the discrete Hausdorff distance $\HD(P, Q)$. 
Perhaps surprisingly, computing the discrete Hausdorff distance is somewhat more complicated than computing the discrete \frechet distance.
The intuition behind this, is as follows.
Because $P$ is $c$-packed, we can upper bound for any decision variable $\rho$ the number of zeroes in the free-space matrix. 
For the discrete \frechet distance, we are interested in a connected walk through the matrix that only consists of zeroes, hence we can find such a walk using depth-first search. 
For the decision variant of the Hausdorff distance, we require instead that the free space matrix contains a zero in every row and in every column. Since this does not have the same structure as a connected path, we require more work and more time to verify this. 
To this end, we assume that in the preprocessing phase we may compute a Voronoi diagram in $O(f(n))$ time, and that the Voronoi diagram has $O(g(n))$ query time. 
In $\R^d$ under the $L_1, L_2, L_\infty$ metric, a Voronoi diagram on $n$ points can be computed in $f(n) = O(n \log n + n^{\lfloor d/2 \rfloor })$ time: it subsequently has $g(n) = O(\log n)$ query time.
In a graph under the shortest path metric a Voronoi diagram on $n$ points can be computed in $O(|G| \log |G|)$ time and it has $O(1)$ query time~\cite{driemel2023discrete}.

Before we go into the details, we make one brief remark: for ease of exposition, 
throughout this section, we assume that our distance oracle $\dist{\alpha}$ is an exact distance oracle.
The reasoning behind this is that we want to simplify our analysis to focus on illustrating the difference between computing the Hausdorff and discrete \frechet distance. Approximate distance oracles may be assumed by introducing $\rho^* = (1 + \frac{\eps}{2}) \rho$ and $\mu = \frac{\eps \rho}{6}$ in the exact way as in Section~\ref{sec:frechet}.
Again our approach immediately also works for subcurve query variants. 

\subparagraph{Hausdorff matrix.}
For values $\rho^* \geq 0$ and some $0 < \eps < 1$ we choose $\mu = \eps \rho^*$.
We define $H^{P^\mu \times Q}_{\rho}$ as a $(0,1)$-matrix of dimensions $|P^\mu|$ by $|Q|$ where rows correspond to the ordered vertices in $Q$ and columns to the ordered vertices in $P^\mu$. 
Let $p_i \in P$ and $p_s$ be its predecessor in $P^\mu$. 
For any $q_j \in Q$, we set the cell corresponding to $(p_i, q_j)$ to zero if there exists a $p' \in P[i, s]$ such that $d(p', q_j) \leq \rho^*$.
Otherwise, we set the cell to one. This defines $H^{P^\mu \times Q}_{\rho^*}$.

\begin{lemma}
\label{lem:hausdorff}
Let $\X$ be any geodesic metric space and $P = (p_1, \ldots, p_n)$ and $Q = (q_1, \ldots, q_m)$ be curves in $\X$.
For any $\rho^* \geq 0$, any $0 < \eps < 1$ and $\mu \leq \eps \rho^*$ :
If there exists a zero in each row and each column of $H^{P^\mu \times Q}_{\rho^*}$ then $\HD(P, Q) \leq (1 + \eps) \rho^*$.
 Otherwise, $H^{P^\mu \times Q}_{\rho^*}$  $\HD(P, Q) > \rho^*$.
\end{lemma}

\begin{proof}
Suppose that there exists a zero in each row and each column of $H^{P^\mu \times Q}_{\rho^*}$. 
Consider a column in $H^{P^\mu \times Q}_{\rho^*}$ corresponding to a vertex $p_s \in P^\mu$ and let its $j$'th row contain a zero. Then for $p_s$ and all $p'$ in between $p_s$ and its successor: $d(p', q_j) \leq \rho^* + \mu \leq (1 + \eps)\rho^*$. Thus $\HD(P \rightarrow Q) \leq (1 + \eps) \rho^*$. 
Similarly, consider the $j$'th row in $H^{P^\mu \times Q}_{\rho^*}$ and let it contain a zero in its column corresponding to $p_j$. Then $d(p_i, q_j) \leq (1 + \eps) \rho^*$ and thus $\FDd(Q \rightarrow P) \leq \rho^*$. 
\end{proof}

From this point on, we observe that the proof of Lemma~\ref{lem:zeroes_per_row} immediately applies to the matrix $H^{P^\mu \times Q}_{\rho^*}$ (indeed, for any zero in the $j$'th row we still obtain a point in the ball $B_1$ -- where the point may be some $p'$ succeeding $p_i$. The segment $S_i$ subsequently must be entirely contained in $B_2$ and so the proof follows). Thus, we conclude:

\begin{corollary}
\label{cor:zeroesH}
Let $P = (p_1, \ldots, p_n)$ be a $c$-packed curve in $\X$.
For any $\rho^* > 0$ and any $k > 1$, denote by $P^\mu$ its $\mu$-simplified curve for $\mu \geq \frac{\eps \rho^*}{k}$. 
For any curve $Q = (q_1, \ldots, q_m) \subset \X$ the matrix $H^{P^{\mu} \times Q}_{\rho^*}$ contains at most  $8 \cdot \frac{c \cdot k}{\eps} \cdot m $ zeroes. 
\end{corollary}

\subsection{Answering A-decision\texorpdfstring{$(Q, \eps, \rho^*)$}{(Q, e, r)}.}

\begin{restatable}{theorem}{HausdorffDecision}
\label{thm:hausdorffdecision}
Let $\X$ be any geodesic metric space and  $P = (p_1, \ldots, p_n)$ be any $c$-packed curve in $\X$. 
Suppose that we can construct a Voronoi diagram on $P$ in $O(f(n))$ time using $O(s(n))$ space and that it has $O(g(n))$ query time. 
We can store $P$ using $O( (n+ s(n)) \log n)$ space and $O((n + f(n)) \log n)$ preprocessing time, such that for any curve $Q = (q_1, \ldots, q_m)$ in $\X$, any $\rho^* \geq 0$ and $0 < \eps < 1$, we can answer A-Decision$(Q, \eps, \rho)$ for the Hausdorff distance in:
\[O \left( \frac{c \cdot m}{\eps} \cdot \log n \cdot  g(n)  \right)  \textnormal{ time}.
\]
\end{restatable}

\begin{proof}
At preprocessing, we construct the data structure of Lemma~\ref{lem:rayshooting}. 
In addition, we create a balanced binary decomposition of $P$ where we recursively split $P$ into two curves $P_1$ and $P_2$ with a roughly equal number of vertices.
For each node in the balanced binary decomposition, we create a Voronoi diagram on the corresponding (sub)curve. 
This takes $O( (n + f(n))\log n)$ total time and $O( (n + s(n)) \log n)$ total space.

At query time, we receive $Q$ and a value $\rho^* \geq 0$. 
We set $\mu = \eps \rho^*$ and in $O( \frac{c \cdot m}{\eps} \log n)$ time we compute the first $8 \frac{c \cdot m}{\eps}$ vertices of $P^\mu$. 

If we discover that $P^\mu$ has more than $8 \frac{c \cdot m}{\eps}$ elements, we report that $\HD(P, Q) > \rho$. 
Indeed: if $\HD(P, Q) \leq \rho$ then each column in $H^{P^{\mu} \times Q}_{\rho^*}$ must contain a zero.
Since there are at most $m$ rows, then by the pigeonhole principle $\HD(P, Q) \leq \rho$ only if $P^\mu$ contains at most  $8 \frac{c \cdot m}{\eps}$ vertices.
We store $P^\mu$ in a balanced binary tree $T$ (ordered along $P^\mu$) in $O(|P^\mu|) = O(\frac{c \cdot m}{\eps})$ time.
We now iterate over each of the vertices in $Q$ and do the following subroutine: 

\subparagraph{Subroutine$(q_j \in Q, T)$:}
We execute each subroutine one by one, passing $T$ along. 
We maintain a stack of pairs of indices.

If after all subroutines $T$ is not empty, we output $\HD(P, Q) > \rho$.

Otherwise, we output $\HD(P, Q) \leq (1 + \eps) \rho$.
\begin{enumerate}
    \item Push the pair $(1, n)$ onto a stack.
    \item Whilst the stack is not empty, do:
    \begin{itemize}
        \item Set $(i, j) \gets \textsc{pop}$ and obtain $P[i, j]$ as $O(\log n)$ roots in our hierarchical decomposition.
        \item For each of the roots, query their Voronoi diagram with $q_j$ in $O(\log n \cdot V_{\eps / 6})$ total time.
        \item Let $p' \in P[i, j]$ be the vertex realising the minimal distance to $q_j$. 
        \begin{itemize}
            \item If $d(p', q) > \rho$: continue the loop with the next stack item.
        \end{itemize}
        \item Otherwise, find its predecessor $p_s$ in $P^\mu$ in $O(\log n)$ time.
        \item Set the cell corresponding to $(p_s, q_j)$ in $H^{P^{\mu} \times Q}_{\rho^*}$ to zero. 
        \item Remove $p_s$ from $T$ in $O(\log \frac{c \cdot m}{\eps} ) = O(\log n)$ time.
        \item Let $p_l$ be the successor of $p_s$ in $P^\mu$.
        \item Push $(i, s)$ and $(l, j)$ onto the stack.
    \end{itemize}
    \item If only $(1, n)$ was ever pushed onto the stack, terminate and output $\HD(P, Q) > \rho$
    
    (no further subroutines are required).
\end{enumerate}

\subparagraph{Runtime.}
Each time we push two items onto the stack, the subroutine has found in the $j$'th row of $H^{P^{\mu} \times Q}_{\rho^*}$ a new cell that has a zero. Thus, the subroutine can push at most $O(\frac{c}{\eps})$ items on the stack.
Per item on the stack, we take $O(\log n \cdot g(n))$ time. Thus, our algorithm takes $O(\frac{c \cdot m}{\eps} \cdot \log n \cdot g(n) )$ total time. 

\subparagraph{Correctness.}
Finally, we show that our algorithm is correct through a case distinction on when we output an answer. 
Let us output our answer after line 3 of the subroutine. 
Then we have found a vertex $q_j \in Q$ whose row in $H^{P^{\mu} \times Q}_{\rho^*}$  contains no zeroes and by Lemma~\ref{lem:hausdorff}: $\HD(P, Q) > \rho$.

Let us output an answer because $T$ contains some vertex $p_s \in P^\mu$ after all the subroutines. 
Then we have found a vertex $p_s \in P^\mu$ whose column in $H^{P^{\mu} \times Q}_{\rho^*}$ contains no zeroes and by Lemma~\ref{lem:hausdorff}: $\HD(P, Q) > \rho$.
Indeed: suppose for the sake of contradiction that the corresponding column contains a zero in row $j$.
Let $p_l$ be the successor of $p_s$ on $P^\mu$.
Because there is a zero in $H^{P^{\mu} \times Q}_{\rho^*}[s, j]$ it must be that there exists a vertex $p_i \in P[s, l - 1]$ with $d(p', q_j) \leq \rho$. 
During Subroutine$(q_j \in Q, T)$, we always have at least one interval $(a, b)$ on the stack where $i\in [a, b]$.
Each time such $(a, b)$ gets found, we find a zero in the $j$'th row and push a pair $(c, d)$ on the stack with $i \in [c, d]$. 
Thus, we eventually must find $p_i$. However, when we find $p_i$ we remove $p_s$ from $T$ which is a contradiction. 

Finally, if none of the above two conditions hold it must be that every row and every column in $H^{P^{\mu} \times Q}_{\rho^*}$ contains a zero and by Lemma~\ref{lem:hausdorff}: $\HD(P, Q) \leq  (1 + \eps)\rho$.
\end{proof}

\subsection{Answering \texorpdfstring{A-value$(Q, \eps)$}{A-value(Q, e)}}

\begin{restatable}{theorem}{HausdorffComputation}
\label{thm:hausdorffcomputation}
Let $\X$ be any geodesic metric space and  $P = (p_1, \ldots, p_n)$ be any $c$-packed curve in $\X$. 
Suppose that we can construct a Voronoi diagram on $P$ in $O(f(n))$ time using $O(s(n))$ space and that it has $O(g(n))$ query time. 
We can store $P$ using $O( (n+ s(n)) \log n)$ space and $O((n + f(n)) \log n)$ preprocessing time, such that for any curve $Q = (q_1, \ldots, q_m)$ in $\X$, and $0 < \eps < 1$, we can answer the subcurve query A-Value$(Q, \eps)$ for the Hausdorff distance in
\[O \left( \frac{c \cdot m}{\eps} \cdot  g(n) \cdot \log^2 n \right)  \textnormal{ time}.
\]
\end{restatable}

\begin{proof}
This proof is simpler than in Section~\ref{sec:frechet}, since we have access to a Voronoi diagram.
During preprocessing, we construct the data structures used in Theorem~\ref{thm:hausdorffdecision}.
Secondly, we note that constructing a TADD on $P$ with itself is dominated by Voronoi diagram construction time.
Thus, we compute a TADD on $P$ with itself in $O(f(n))$ time.
This TADD is stored as a set of sorted intervals $\mathcal{I} = \{ I_s = [c_s, 2 \cdot c_s] \}$.
We create the set $\mathcal{H} = \{ H_s = [ \frac{1}{2} c_s, 4 \cdot c_s] \}$.
Note that $\mathcal{H}$ is a set of intervals which may overlap. We keep $\mathcal{H}$ sorted by $c_s$.

Given a query $(Q, \eps)$ we first do the following: 
we compute the Hausdorff distance from $Q$ to $P$ in $O(m \cdot g(n))$ time (by querying for every point in $Q$, the Voronoi diagram in $P$).
Let this Hausdorff distance be $\lambda$, we create an interval $H^* = [\lambda, 2 \lambda]$.

We now claim, that $\rho = \HD(P, Q)$ is contained in either an interval in $\mathcal{H}$ or in $H^*$.
Indeed, consider for each $q \in Q$ a disk with radius $\rho$ centered at $q_j$.
Each of these disks must contain at least one point of $P$.
Moreover, there must exist at least one $q_j \in Q$ which has a point $p_i$ on its border (else, the Hausdorff distance is smaller than $\rho$). 
Consider the disk $D_j$ centered at $q_j$ with radius $\frac{1}{2} \rho$. There exist two cases:
\begin{enumerate}[a)]
    \item $D_j$ contains a point $p_l \in P$. 
    Then $\rho \in [\frac{1}{2} \cdot d(p_l, p_i), 4 \cdot d(p_l, p_i)]$ and so $\rho$ is in $\mathcal{H}$.
    \item $D_j$ contains no points in $P$. 
    Then $\HD(Q \to P)$ is at least $\frac{1}{2} \rho$ (and at most $\rho$) so $\rho \in H^*$. 
\end{enumerate}

\noindent
Having observed this, we simply do a binary search over $\mathcal{H}$ (and check $H^*$ separately). 
For each interval $H_s = [ \frac{1}{2} c_s, 4 \cdot c_s]$, we query: A-Decision$(Q, \eps, 4 \cdot c_s)$. 
We choose $\rho^* = 4 \cdot c_s$ and $\mu = \eps  \frac{1}{2} c_s$.
By Corollary~\ref{cor:zeroesH}, the matrix $H_{\rho^*}^{P^\mu \times Q}$ contains $O(\frac{c \cdot m}{\eps})$ zeroes, which we identify with the algorithm from Theorem~\ref{thm:hausdorffdecision} in $O(\frac{c \cdot m}{\eps} \cdot g(n) \cdot \log^2 n)$ time. 
If $s$ is the smallest integer $s$ for which the matrix $H_{\rho^*}^{P^\mu \times Q}$ contains a zero in every row and every column, then we obtain a $(1+\eps)$-approximation of the Hausdorff distance. We obtain $s$ by inspecting all zeroes in the matrix $H_{\rho^*}^{P^\mu \times Q}$, and computing the minimal required pointwise distance. 
\end{proof}

\bibliography{refs}

\end{document}